%% file: oort_cloud_transfer.tex
\documentclass[a4paper,fleqn,usenatbib]{mnras}

\usepackage{newtxtext,newtxmath}
\usepackage[T1]{fontenc}
\usepackage{ae,aecompl}
\usepackage{graphicx}
\usepackage{amsmath}
\usepackage{amssymb}

\newcommand\kms{\ensuremath{\mathrm{km\,s}^{-1}}}	
\newcommand\e{\ensuremath{_{\mathrm{\earth}}}}  
\newcommand\mstar{\ensuremath{M_{\star}}}   
\newcommand\msun{\ensuremath{\mathrm{M}_{\odot}}} 
\newcommand\noc{\ensuremath{N_{\mathrm{OC}}}} 
\newcommand\oc{\ensuremath{{\mathrm{OC}}}} 

\defcitealias{2017MNRAS.464.2290M}{MB17}

\title[Capture of exocomets and the erosion of the OC]{Capture of exocomets and the erosion of the Oort cloud due to stellar encounters in the Galaxy}

\author[J. Hanse et al.]{
J.~Hanse,$^{1}$\thanks{E-mail:\,\,hanse, jilkova\,\,@strw.leidenuniv.nl}
L.~J\'{i}lkov\'{a},$^{1}$\footnotemark[1]
S.\,F.~Portegies Zwart$^{1}$
and F.\,I.~Pelupessy$^{2,1}$
\\
$^{1}$Leiden Observatory, Niels Bohrweg 2, Leiden, 2333\,CA, The Netherlands \\
$^{2}$Netherlands eScience Center, Science Park 140, 1098 XG Amsterdam, The Netherlands
}

\date{Accepted XXX. Received YYY; in original form ZZZ}
\pubyear{2017}

\begin{document}
\label{firstpage}
\pagerange{\pageref{firstpage}--\pageref{lastpage}}
\maketitle

\begin{abstract}
The Oort cloud (OC) probably formed more than 4\,Gyr ago and has been moving with the Sun in the Galaxy since, exposed to external influences, most prominently to the Galactic tide and passing field stars. 
Theories suggest that other stars might posses exocomets distributed similarly to our OC. 
We study the erosion of the OC and the possibility for capturing exocomets during the encounters with such field stars. 
We carry out simulations of flybys, where both stars are surrounded by a cloud of comets. 
We measure how many exocomets are transferred to the OC, how many OC's comets are lost, and how this depends on the other star's mass, velocity and impact parameter. 
Exocomets are transferred to the OC only during relatively slow ($\lesssim0.5$\,km\,s$^{-1}$) and close ($\lesssim10^5$\,AU) flybys and these are expected to be extremely rare. 
Assuming that all passing stars are surrounded by a cloud of exocomets, we derive that the fraction of exocomets in the OC has been about $10^{-5}$--$10^{-4}$. 
Finally we simulate the OC for the whole lifetime of the Sun, taking into account the encounters and the tidal effects. 
The OC has lost 25--65\% of its mass, mainly due to stellar encounters, and at most 10\% (and usually much less) of its mass can be captured. However, exocomets are often lost shortly after the encounter that delivers them, due to the Galactic tide and consecutive encounters.

\end{abstract}

\begin{keywords}
Oort Cloud -- comets: general -- celestial mechanics
\end{keywords}


\section{Introduction}
\label{sec:intro}

For several decades, it has been argued that the Solar system is surrounded by $10^{10}$--$10^{12}$ icy bodies, called the Oort cloud \citep[OC,][]{1950BAN....11...91O}.
This cloud is the source of the observed long-period comets (LPCs, with orbital period $>200$\,yr) which typically have large semi-major axes of about $10^4$\,AU. 
The current picture describes the OC as an isotropic distribution of objects orbiting outside the planetary region (perihelia\,$\gtrsim32$\,AU) with large semi-major axes of about $2\times10^3$--$10^5$\,AU \citep{2015SSRv..197..191D}.
The orbits of OC objects can be perturbed by the Galactic tide \citep[established by][]{1986Icar...65...13H}, flybys of passing stars \citep[pioneered by][]{1950BAN....11...91O}, and encounters with giant molecular clouds \citep[e.g.][]{1985AJ.....90.1548H,2009CoSka..39...85J}.
These may shift the OC objects perihelia to the planetary region.
There they receive energy kicks from planets which can cause them to escape the Solar system or move to the outer parts of the OC, where they are subject to stronger external perturbations \citep{2009Sci...325.1234K}.
Some OC objects end up on orbits with perihelia sufficiently small ($\lesssim5$\,AU) to be observable.

\subsection{Oort Cloud formation}
Theoretical studies argue that the OC formed in about the first $0.5$\,Gyr of the Solar system's evolution.
After that it has been constantly and gradually depleted by the external perturbations mentioned above \citep[see][for a more detailed review of the OC formation]{2015SSRv..197..191D}.
Even though the details are still under discussion, an interplay between planetary scattering of comets from the early Solar system and external influences was identified as a key mechanism for the delivery of comets to the OC. 
The external influences include the Sun's birth cluster and embedding gas, and the Galaxy itself.
First, the comets were scattered by close encounters with planets into orbits with large semi-major axes.
Then their perihelia were further lifted beyond the planetary region by influences from outside the Solar system.

\citet{2000Icar..145..580F}, \citet{2006Icar..184...59B}, and \citet{2008Icar..197..221K} studied early OC formation in the Sun's birth cluster.
In this formation scenario early in the history of the Solar system, comets are scattered from the disk during planetary formation.
Then their orbits evolve under influence of the stars that formed in the same molecular cloud (Solar siblings, \citealp{2009ApJ...696L..13P}) and the gas they are embedded in.
However, \citet{2007Icar..191..413B} pointed out the effect of gas drag in the primordial Solar nebula which circularizes the orbits of comets before they can be scattered and prevents their deposition in the OC.

\citet{2013Icar..225...40B} studied delayed OC formation, after the Sun left its birth cluster.
Comets are scattered out of the planetary region during the late dynamical instability of the Solar system (about $0.5$\,Gyr after its formation) according to the Nice model \citep{2005Natur.435..459T}.
Here the external influences are the overall Galactic tide and flybys with field stars \citep[e.g.][]{1993ASPC...36..335T,2004ASPC..323..371D,2008CeMDA.100....1B}.
Some support for the delayed OC formation also comes from the results of \citet{2017arXiv170403341N}.
They showed that if the OC formed while the Sun resided in its birth cluster, the outer parts ($\gtrsim 3000$\,AU) of the OC would be stripped due to the cluster gravitational potential and flybys with Solar siblings.

\subsection{Exo-Oort Clouds}
Similarly to the Solar system, the formation of possible OC analogues around other stars (exo-OCs) involves planetary scattering of comets from the disk and detachment of their orbits by external influences \citep[e.g.][]{1993ASPC...36..335T,2017MNRAS.464.3385W}.
\citet{2013MNRAS.429L..99R} simulated the effect of planet--planet scattering on planetesimal disks.
They showed that an isotropic cloud with a typical size of 100--1000\,AU can be formed if one the planets is ejected from the system.
These clouds resemble the Solar system's scattered disk, but with an isotropic distribution due to stronger interactions during the planet--planet scattering phase.

Despite several attempts (e.g. \citealp{1991Icar...91...65S} or \citealp{2010AAS...21560106B}), comets around other stars have never been directly observed \citep{2015SSRv..197..191D}.
However, some features of debris disks observations have been interpreted as providing indirect evidence.
In particular, short-term variations in gas absorption has been detected in several debris disks (see \citealp{2015AdAst2015E..26W} for a summary) and are believed to result from the evaporation of active minor bodies on eccentric star-grazing orbits \citep{1990A&A...236..202B}.

\subsection{Exocomets in Solar system}
\citet{1987Icar...69..185S} was among the first to investigate the possible presence of exocomets in the Solar system.
He estimated the impact rate of exocomets on the terrestrial planets during the Solar system's passage through an exo-OC.
He concluded these to be relatively rare, about 2--10 impacts of exocomets on all terrestrial planets together over the lifetime of the Solar system.
This however still easily dominates the number of impacts of free interstellar comets (which are not bound to any stellar system).

\citet[][]{2010Sci...329..187L} studied the possibility of exchange of comets between Solar siblings while still in their birth cluster.
They concluded that over $90$\% of OC objects might be of extrasolar origin.
They assumed however, that the clouds were formed early in the history of the Solar system, which is in contradiction with \citet[][see above]{2007Icar..191..413B}.

\subsection{Solar sojourn through the Galaxy}
The location of the Sun in the Galaxy affects the frequency and characteristics of encounters with field stars \citep[][hereafter \citetalias{2017MNRAS.464.2290M}]{2017MNRAS.464.2290M}.
It also sets the strength of the Galactic tide that influences OC formation \citep{2010A&A...516A..72B,2011Icar..215..491K} and continuously affects its evolution (e.g. \citealp{1986Icar...65...13H}, \citealp{2007AJ....134.1693H}).
The Galactocentric radius at which the Sun was born and its past orbit in the Galaxy are under discussion.
It is possible that the Sun has migrated up to a few kpc in the Galactic disk \citep[][]{2008ApJ...684L..79R,2015MNRAS.446..823M}.

\subsection{This work}

Motivated by the possible existence of exo-OCs and the recent work on flybys with field stars along Solar sojourn through the Galaxy \citepalias{2017MNRAS.464.2290M}, we study their effect on comets and exocomets in the clouds.
We focus on the possibility that the Sun's OC captures exocomets from the field stars encountered by the Sun.
First, we estimate the transfer efficiency of encounters while varying their properties (mass of the star, impact parameter and impact velocity).
We compare these to the encounters expected to occur along the Sun's orbit.
Second, we perform simulations of the Sun and its OC as they orbit through the Galactic potential, while being exposed to a realistic sequence of of field stars flybys, each with an exo-OC, and the Galactic tide. 
We consider three Solar orbits assuming different radial migration:
starting in the inner disk and moving outward, starting in the outer disk and moving inward, and an orbit without substantial migration.

\section{Methods}
\label{sec:methods}

For our simulations we use the Astronomical Multipurpose Software Environment, or \textsc{Amuse} \citep{2013CoPhC.184..456P,2013A&A...557A..84P}\footnote{\textsc{Amuse} is an open source software framework available at \texttt{https://github.com/amusecode/amuse}.}.
We carry out two types of simulations. 
First, we perform a parameter space study and investigate the fraction of the OC that is lost and captured in individual encounters.
We vary the encounter parameters in ranges in which we expect the Sun's OC to capture exocomets (Secs.~\ref{sec:gridsim} and \ref{sec:resultsGridsim}).
Second, we integrate the complete orbit of the Sun\footnote{By \emph{orbit of the Sun} or \emph{Solar orbit} we refer to the complete trajectory of the Sun in the Galaxy since its formation $\sim4.5$\,Gyr ago. We do not take the effect of the Sun's birth cluster into account.} together with its OC through the Galaxy while taking the effect of encounters with field stars into account (Secs.~\ref{sec:orbitsim} and \ref{sec:resultsOrbitsim}).

\subsection{Structure of Oort Cloud and exo-OCs}
\label{sec:oc}

We refer to bodies in the OC as comets (although some of them might actually be asteroids \citealp{2015MNRAS.446.2059S} or some other kinds of objects). 
We assume the comets initially follow a spherically symmetric and isotropic distribution, following \citet{2008CeMDA.102..111R}.
Their initial semi-major axes, $a$, have values of $3\times10^3$--$10^5$\,AU distributed over a probability density proportional to $a^{-1.5}$.
Initial eccentricities, $e$, are chosen with a probability density distribution $\propto e$ where we pick only the values resulting in perihelia outside the planetary region, $q>32$\,AU.
We generate additional initial orbital elements of the comets such that the cosine of inclination, argument of perihelion, longitude of the ascending node, and mean anomaly have uniform distributions.
The initial radial density profile of the OC is then $\propto r^{-3.5}$, where $r$ is the distance between the comets and the Sun \citep{1987AJ.....94.1330D}.
This model is commonly used to approximate a thermalized OC \citep{2002A&A...396..283D,2008CeMDA.102..111R,2011A&A...535A..86F,2014Icar..231..110F,2014MNRAS.442.3653F}.

Estimates of the number of OC comets larger then $2.3$\,km range from $10^{10}$ to $10^{12}$ \citep[][]{1996ASPC..107..265W,2013Icar..225...40B}.
The total mass of the OC, $M_\oc$, is also uncertain.
For example, \citet{2005ApJ...635.1348F} estimates $M_\oc=2$\,--\,$40$\,M$\e$.
Assuming $10^{11}$\,comets with a total mass of 10\,M$\e$ and a typical cometary velocity of about $200$\,m\,s$^{-1}$ (the circular velocity at $2\times10^4$\,AU from the Sun), the two-body relaxation time-scale is of the order $10^{17}$\,yr.
Two-body interaction of the comets can therefore be neglected.
The comets also have a much smaller mass than the Sun and the other stars.
Therefore we represent the comets in our simulations with \noc{} zero-mass test particles.

We assume that all field stars encountering the Sun posses an exo-OC.
The density profiles of exo-OCs are assumed similar to the Solar OC, but scaled with the mass of their parent stars.
The number of test particles in an exo-OC scales as $N_{\mathrm{exoOC}} = \noc{}(\mstar{}/\msun{})$, where \mstar{} is the mass of the exo-OC's parent star.
We assume that the outer edge of the exo-OC is proportional to the Hill radius of the star in the Galactic potential.
The inner edge of the exo-OC is influenced by the planets in the system and the gravitational tide of the surrounding environment.
There is no obvious way to predict the size of the other star's planetary systems and the inner edge of its exo-OC.
For simplicity, we assume that the planetary system is similar to the Solar system and the exo-OC inner edge is also proportional to its parent star's the Hill radius.
We obtain the inner and outer semi-major axes of the exo-OC by scaling the inner and outer semi-major axes of the Sun's OC by a factor
\begin{equation}
\label{eq:scaling_ocs}
\frac{R_{\star}}{R_{0\odot}}
\left(\frac{\mstar{}}{\mathrm{M}_{\sun}} \frac{M_{\mathrm{G}}(R_{0\odot})}{M_{\mathrm{G}}(R_{\star})}\right)^{1/3}.
\end{equation}
Here $R$ is the cylindrical Galactocentric radius, $R_{0\odot}=8.5$\,kpc is the current Galactocentric radius of the Sun, and $R_{\star}$ is the Galactocentric radius of the star.
$M_{\mathrm{G}}(R)$ is the cumulative mass of the Galaxy enclosed within radius $R$ and is calculated consistently with the adopted Galactic potential (see Sec.~\ref{sec:gal_pot}).
Same as for the Solar system, the minimal pericenter distance of the exocomets is set to $32$\,AU.

We integrate the motion of the two stars and their OCs using the $N$-body code \textsc{huayno} \citep{2012NewA...17..711P}.
We use the {\it HOLD drift-kick-drift} integrator method with a timestep corresponding to $0.03$ of the inter-particle free-fall time.

\subsection{Galactic potential and \textsc{bridge}}
\label{sec:gal_pot}

We use a gravitational potential that includes the effects of the bar and spiral arms.
We follow the model described by \citetalias{2017MNRAS.464.2290M} (an updated version of \citealp{2015MNRAS.446..823M}) in which analytical prescriptions for the Galactic bar, spiral arms and axisymmetric background potential are combined.
We use the same Galactic model as \citetalias{2017MNRAS.464.2290M}, who calculated the rate of stellar encounters along migrating Solar orbits (see Sec.~\ref{sec:enc}).
In this model, the axisymmetric background potential of \citet{1991RMxAA..22..255A} is used.
It consist of a bulge \citep{1911MNRAS..71..460P}, disc \citep{1975PASJ...27..533M} and a logarithmic dark matter halo.
The Galactic bar is represented by a three-dimensional Ferrers potential \citep[e.g.][]{1987gady.book.....B}.
It rotates with an angular velocity of $55\,$km\,s$^{-1}$\,kpc$^{-1}$ and its mass is about $70$\% of the bulge total mass of $1.41\times10^{10}\,$M$_{\odot}$.
The spiral structure has two arms that are modelled by the three-dimensional potential of \citet{2002ApJS..142..261C} and rotates with angular velocity $25\,$km\,s$^{-1}$\,kpc$^{-1}$.

The enclosed mass $M_{\mathrm{G}}(R)$ in Eq.~\ref{eq:scaling_ocs} is calculated using an analytic prescription for the axisymmetric potential that is unperturbed by the bar and spiral arms (these components represent a redistribution of the mass of the bulge and the disk).

The Galactic potential is coupled to the system of stars and their comets using \textsc{bridge} (\citealp{2007PASJ...59.1095F}, \citealp{amuse_book}), which is an extension of the mixed variable symplectic scheme \citep{1991AJ....102.1528W}.
To account for the rotating components of the potential, we integrate the equations of motion in a non-inertial rotating reference frame (so-called \textsc{rotating bridge}, \citealp{pelupessy_prep_2017}, \citealp{amuse_book}).
We use a \textsc{bridge} timestep of $\leq1$\,Myr
\footnote{The \textsc{bridge} timestep is adjusted such that it is $\leq1$\,Myr and an exact integer number of \textsc{bridge} steps equals the integration interval. This is to make sure that the \textsc{bridge} integration runs exactly until the end of the integration interval.}
 and a sixth-order integrator, which takes 13 substeps ranging from $0.12$ to $0.4$\,Myr each.

The computation time scales linearly with the total number of particles.
To balance the computation time and resolution, we use different $N_{\oc}$ (resulting in different $N_{\mathrm{exoOC}}$) for the parameter space study (see Sec.~\ref{sec:gridsim}).
For full orbit simulations, we use $N_{\oc}=2\times10^4$ (Sec.~\ref{sec:orbitsim}).
Each of these particles then represents about $5\times10^6$\,actual OC comets, assuming their total number in the OC is $10^{11}$.

Because the comets are represented by test particles, the integration represents an embarrassingly parallel problem.
That allows us to distribute the calculations across many CPUs.
Each of those integrates the motion of two stars in the Galactic potential together with a different part of the clouds.
The simulations were performed on desktop computers and the Para Cluster at the Leiden Observatory.
We used up to $10$\,desktops computers for the full orbit simulations.
Each orbit took about $20$ to $40$ days of CPU time, depending on the encounters parameters and the properties of the CPUs, which translates to two to four days of distributed computing.
The parameter space study consists of $288$ individual encounter simulations (three bins in mass, eight in impact parameter, and $12$ in velocity, Sec.~\ref{sec:gridsim}) taking between a few minutes and $14$\,hours of CPU time per encounter.

\subsection{Parameter space study of single encounters}
\label{sec:gridsim}

\begin{table}
\caption{Number of test particles representing the Sun's OC ($N_{\oc}$) and the exo-OC of the field star ($N_{\mathrm{exoOC}}$) in the parameter space study of single encounter (Sec.~\ref{sec:gridsim}) for different \mstar{}.
The total number of test particles ($N_{\oc}+N_{\mathrm{exoOC}}$) is of the order of $10^5$ for all three \mstar{}.
}
\label{tab:n_mdv}
\center
\begin{tabular}{rrr}
\hline
$\mstar{}\,[\msun{}]$ & $N_{\oc}$ & $N_{\mathrm{exoOC}}$ \\
0.3     &   182575  &  54773  \\
2.7     &   60859   &  164317 \\
24.1    &   20371   &  490918 \\
\hline
\end{tabular}
\end{table}

In the first series of simulations, we investigate what portions of the Sun's OC is lost and captured from a passing star in an individual encounter.
We simulate encounters of the Sun with another star where both stars are surrounded by cometary clouds with initial conditions described in Sec.~\ref{sec:oc} with and without including the effect of the Galactic tide.
Each encounter is characterized by the mass of the passing star, \mstar{}, the impact parameter vector, $\mathbfit{d}$, and the velocity vector at infinity with respect to the Sun, $\mathbfit{v}_{\star}$.
We consider the following values of the mass of the field star: $M_{\star}=0.3,\,2.7,\,24.1$\,\msun{}.
These values are chosen such that they sample the whole range of stellar masses considered in the complete orbit simulation ($0.082$--$60$\,\msun{}, Sec.~\ref{sec:orbitsim}).
For the impact parameter we take $\log (d/\mathrm{AU})=2.25$\,--\,$5.75$, with a step of $0.5$.
For the velocity we take $\log \left(v_{\star}/(\mathrm{km\,s}^{-1})\right) = -1.125$\,--\,$1.875$ with a step of $0.25$.
Here $d$ and $v_{\star}$ are the magnitudes of the vectors $\mathbfit{d}$ and $\mathbfit{v}_{\star}$.

The stars are $8$\,pc apart at the start of the simulation and we stop the simulation when their separation is again $8$\,pc.
As mentioned earlier, we run simulations both with and without including the Galactic tide effects.
In the case without a Galactic potential, the direction of $\mathbfit{d}$ and $\mathbfit{v}_{\star}$ are not important because both clouds are spherically symmetric, otherwise the directions of $\mathbfit{d}$ and $\mathbfit{v}_{\star}$ are determined as described in Sec.~\ref{sec:enc_start}.

We assume the encounters to happen at the current Solar position.
The inner and outer radii of the exo-OCs are scaled using Eq.~\ref{eq:scaling_ocs}, which reduces to a factor of $(\mstar{}/\msun{})^{1/3}$ because $R_{\star} \simeq R_{0\odot}$.
We represent the OC and the exo-OC by a different number of test particles for each  \mstar{} such that the total number of test particles ($N_{\oc}+N_{\mathrm{exoOC}}$) is of the order of $10^5$.
$N_{\oc}$ is chosen such that the resolution for each \mstar{} is similar and the computation time is reasonable.
The number of test particles in each cloud is summarized in Table~\ref{tab:n_mdv}.

\begin{figure}
\center
\includegraphics[width=0.49\textwidth]{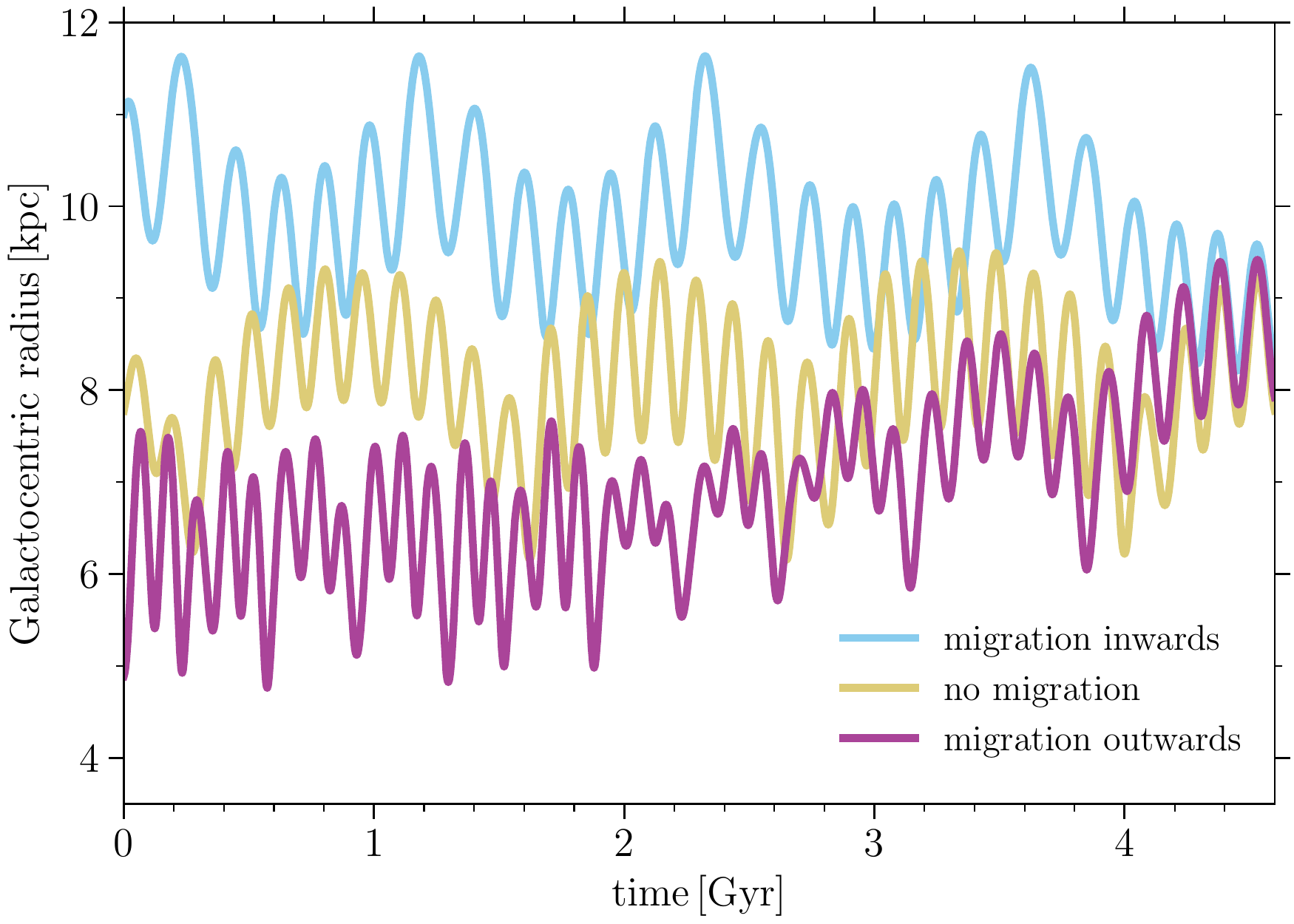}
\caption{Evolution of Solar Galactocentric radius for three orbits assuming different radial migration of the Sun \citepalias{2017MNRAS.464.2290M}.
The orbits were integrated in an analytic model of the Galactic potential including the effects of the bar and spiral arms.}
\label{fig:solar_orbit}
\end{figure}

\subsection{Full orbit simulation}
\label{sec:orbitsim}

In the second series of simulations, we study the evolution of the Sun's OC during its lifetime in the Galactic disk (after escaping from the Sun's birth cluster, see e.g. \citealp{2009ApJ...696L..13P}).
We simulate encounters with field stars surrounded by exo-OCs along the Solar orbit in the Galaxy.
To account for uncertainty in the Solar orbit and the Galactocentric radius at which the Sun was born, we follow \citetalias{2017MNRAS.464.2290M} and consider three orbits with different radial migrations to which we further refer as three {\it types of orbits}.
In Fig.~\ref{fig:solar_orbit} we show the evolution of the Solar Galactocentric radius for these three orbits.
An outward migrating orbit which starts at a Galactocentric radius of about $6$\,kpc, a non migrating orbit which starts at $8$\,kpc, and an inward migrating orbit which starts at $11$\,kpc.
At $t=4.5$\,Gyr, all orbits end within the expected uncertainties of the current position and velocity vectors of the Sun: $\mathbfit{r}_{\odot}=(-8.5,\,0,\,0.02)$\,kpc and $\mathbfit{v}_{\odot}=(-11.1,\, 238.4,\,7.25)$\,km\,s$^{-1}$ \citepalias{2017MNRAS.464.2290M}.
The frequency of stellar encounters depends on the local density and velocity dispersion and is therefore determined by the Solar orbit.

\subsubsection{Encounter sets}
\label{sec:enc}

\citetalias{2017MNRAS.464.2290M} computed the rate of stellar encounters the Sun experienced along the three types of orbits and the properties of these encounters.
Each encounter is specified by the time $t$ it occurs, and by \mstar{}, $d$, and $v_{\star}$ (see Sec.~\ref{sec:gridsim}). 
Their procedure (based on \citealp{2008CeMDA.102..111R} and \citealp{2014MNRAS.442.3653F}) involves randomly picking values from probability distributions of $t$, \mstar{}, $d$, and $v_{\star}$ and takes into account the fact that the probability of an encounter is proportional to its velocity $v_{\star}$.
This is especially important for low velocity encounters which have the strongest effect on the OC but are rare.
We use the same procedure to generate the encounters properties in our simulations for the three types of orbits. 

We also need to determine the direction of the encounters taking the Solar apex motion into account, which is not discussed by \citetalias{2017MNRAS.464.2290M}.
We assume that the Sun's velocity with respect to the Local Standard of Rest (LSR) is constant, and the velocity direction of field stars with respect to the LSR is random. 
This results in an anisotropic distribution of the encounter velocity vectors with respect to the Sun ($\mathbfit{v}_{\star}$).
Their direction  concentrates in the Solar motion antapex.

Using the distributions derived by \citetalias{2017MNRAS.464.2290M}, we generate the encounter parameters along the three orbits.
We set the maximum impact distance of the field star to $5\times10^{5}$\,AU (value of $D$ in equation~6 of \citetalias{2017MNRAS.464.2290M}).
This results in $1.5\times 10^5$ encounters for an inward migrating orbit, $2.7\times 10^5$ encounters for an orbit without radial migration, and $4.4\times 10^5$ encounters for an outward migrating orbit\footnote{Note that these numbers are higher than those given by \citetalias{2017MNRAS.464.2290M} who used a smaller maximum impact distance of $4.5\times10^{5}$\,AU.}.
We call the complete batch of encounters generated along an individual orbit an {\it encounter set}.

We compare the distributions of encounter properties along orbits with different migration types in Appendix~\ref{app:encounters_mig} and Fig.~\ref{fig:contours_mig}.
As shown by \citetalias{2017MNRAS.464.2290M} (their Fig.~4) the distribution of encounter parameters depends only weakly on the Solar migration and most encounters are with low-mass stars ($\mstar{}<1\,\msun{}$) with velocities of $20$--$100\,\kms{}$.
The total number of encounters in an encounter set, however, depends on the Sun's radial migration.

\subsubsection{Selection of the simulated encounters}
\label{sec:enc_selection}

We assume that the Sun encounters one star at a time.
Each encounter starts at a distance of $8$\,pc between the Sun and the other star and continues until they are again separated by $8$\,pc.
Depending on the encounter parameters (most importantly the velocity), the simulation of each encounter takes a different time.
We can simulate only a limited number of encounters during the lifetime of the Sun ($4.5$\,Gyr).
We pick the encounters we simulate from the encounter sets drawn as described in Sec.~\ref{sec:enc} as follows.

We choose to simulate encounters that deliver the highest impulse to the Sun.
We order all encounters in the set by decreasing $M_{\star}/(v_{\star}d)$, where we assume that the relative velocity of the Sun and the star is constant during the encounter and equals $v_{\star}$.
Hence $d$ and $v_{\star}$ also represent distance and velocity at the perihelion of the encounter.
The impulse delivered to the Sun $\propto M_{\star}/(v_{\star}d)$ has been used before as a proxy for encounter strength, as measured by the number of LPCs injected from the OC into orbits with smaller pericenters \citep{2009Sci...325.1234K,2011A&A...535A..86F,2014MNRAS.442.3653F}.
We approximate the duration of an encounter by $t_{\mathrm{enc}}=16\,\mathrm{pc}/v_{\star}$, where we again assume a constant velocity during the encounter.
In the actual simulation, $\mathbfit{v}_{\star}$ is affected by the Galactic tide and the two-body interaction of the Sun and the star.
For encounters with a high velocity, the timescale of the encounter is much shorter than the interaction timescale with the Galactic potential and $t_{\mathrm{enc}}$ represents a good first order estimate to the actual encounter timescale.
For low velocity encounters ($v_{\star}\lesssim 1\,\kms{}$), the timescales of the encounter and of the interaction with the Galactic potential are comparable to each other and the discrepancy is larger (see also Sec.~\ref{sec:enc_start}).
However, such encounters are rare (see Figs.~\ref{fig:mdv} and~\ref{fig:contours_mig}) and the overall inaccuracy due to using $t_{\mathrm{enc}}$ for the selection of the simulated encounters is small.
We select $j$ encounters such that $\sum_{i\leq j}t_{\mathrm{enc},i}<\alpha 4.5$\,Gyr, where $i$ and $j$ are indices ordering the encounters from highest to lowest transferred impulse.
Here $\alpha$ is a small correction factor to make sure that the the integration time of the full orbit is as close as possible to $4.5$\,Gyr.
Encounters selected by this procedure add up to between $4.39$ and $4.61$\,Gyr of the actual simulation times (difference $<3$\% from the desired $4.5$\,Gyr)

The selection typically results in $\sim3900$\,encounters for the orbit without migration (and $\sim3600$ and $4000$, for inward and outward migration, respectively).
The actual number depends on the parameters of individual encounters in the set (Sec.~\ref{sec:enc}).
We simulate these encounters in the order of the time $t$ as given by \citetalias{2017MNRAS.464.2290M}. 
This time $t$ does not generally correspond to the time of the encounter in the simulation, but is only used to order the encounters.
We refer to the selection of encounters as described in this section as an \emph{encounter series}.
The effect of the remaining encounters in the set (i.e. those not selected for an encounter series) is not taken into account.

The number of transferred and lost particles depends on the encounter parameters.
As we discuss in Sec.~\ref{sec:resultsGridsim}, particles are captured from the exo-OCs in encounters that appear only once in about $20$--$40$ encounter sets (depending on the migration), or in other words, only once in about $20$--$40$ Solar lifetimes.
Because of this low probability, we manually pick $30$ encounter sets that include encounters we expect to result in exocomet capture, since their parameters fall within a bin of the parameter space study that has $N_{\mathrm{cap}}>0$, and we simulate their effect on the Sun's OC as it evolves along its orbit.
These simulations give us insight into the evolution of captured exocomets in the Sun's OC under the influence of consecutive stellar encounters and the Galactic tide.
We also simulate six orbits (two per each orbit type) with encounter sets not resulting in exocomet capture.
We discuss the results in Sec.~\ref{sec:resultsOrbitsim}.

\subsubsection{Starting position and velocity of the encountering star}
\label{sec:enc_start}

The velocity of slow encounters with $v_{\star}\lesssim 1\,\kms{}$ is influenced by the Galactic tide.
The distance and velocity vectors at the perihelion of the encounter, $\mathbfit{d}_{\mathrm{peri}}$ and $\mathbfit{v}_{\mathrm{peri}}$, strongly depend on the initial direction of the vectors $\mathbfit{d}$ and $\mathbfit{v}_{\star}$ with respect to the Galactic potential.
Different directions of the vectors $\mathbfit{d}$ and $\mathbfit{v}_{\star}$, can result in very different $d_{\mathrm{peri}}$ and $v_{\mathrm{peri}}$ which in turn results in a different effect on the clouds.
Therefore, to determine the initial position and velocity vectors of the field star, we first calculate $\mathbfit{d}_{\mathrm{peri}}$ and $\mathbfit{v}_{\mathrm{peri}}$ as they result from an isolated two-body problem where only the gravity of the two stars is considered (i.e. without the Galactic tide). 
Note that the direction of $\mathbfit{v}_{\mathrm{peri}}$ depends on the direction of $\mathbfit{v}_{\star}$ which is anisotropic\,---\,see the description in Sec.~\ref{sec:enc}.

We need to calculate the distance and velocity vectors between the Sun and the field star, $\mathbfit{r}$ and $\mathbfit{v}$, at the beginning of the encounter, $t=t_{\mathrm{enc},0}$, that is find $\mathbfit{r}(t_{\mathrm{enc},0})$ and $\mathbfit{v}(t_{\mathrm{enc},0})$, such that $r(t_{\mathrm{enc},0}) = 8$\,pc and at some time in the future $\mathbfit{r}(t) \approx \mathbfit{d}_{\mathrm{peri}}$ and $\mathbfit{v}(t) \approx \mathbfit{v}_{\mathrm{peri}}$.
We use an iterative procedure to calculate $\mathbfit{r}(t_{\mathrm{enc},0})$ and $\mathbfit{v}(t_{\mathrm{enc},0})$.
Starting at the position of the Sun, we integrate the motion of the Sun in the Galactic potential for half of the encounter time-scale $t_0=8\,\mathrm{pc}/v_{\star}$.
Next we insert the field star in the simulation at $\mathbfit{d}_{\mathrm{peri}}$ and $\mathbfit{v}_{\mathrm{peri}}$ from the Sun and integrate the system backwards until the separation between the star and the Sun is $8$\,pc.
The time of the backwards integration, $t_1$, is in general different from $t_0$ and the Sun does not end up at its original starting position on its orbit.
We repeat the procedure with the updated timescale $t_1$: we integrate the motion of the Sun in the Galactic potential from the intial point on its orbit for time $t_1$;
add the field star at $\mathbfit{d}_{\mathrm{peri}}$ and $\mathbfit{v}_{\mathrm{peri}}$ (the same vectors as in the first step), 
and integrate backwards until the separation of the Sun and the star is $8$\,pc, which gives time $t_2$.
We repeat this procedure $n$ times until the encounter time-scale from two consecutive iterations differs by less than 1\%, that is until $|t_n - t_{n-1}|/t_{n-1}<0.01$.
This procedure assures that the initial separation between the stars is $8$\,pc, and that their position and velocity at the encounter perihelion are $\mathbfit{d}_{\mathrm{peri}}$ and $\mathbfit{v}_{\mathrm{peri}}$.
We use these position and velocity vectors of the Sun and the field star as the initial conditions for the encounter calculation.

For some low velocity encounters the above method does not converge.
The encounter takes too long and there is a strong interaction with the Galactic potential.
Such encounter occurs in approximately 1\% of the encounter sets and we describe the procedure to derive its $\mathbfit{r}(t_{\mathrm{enc},0})$ and $\mathbfit{v}(t_{\mathrm{enc},0})$ in Appendix~\ref{app:vel_ini_for_low_v}.

\subsubsection{Position and velocity of the Sun}
\label{sec:sun_orbit_adjust}

Due to gravitational interaction with the field star during the encounter, the position and velocity of the Sun at the end of an encounter is generally not the same as for the original orbit calculated using a smooth analytic Galactic gravitational potential (Sec.~\ref{sec:gal_pot}).
We use orbits that undergo a strong radial migration and a small change in phase space position of the Sun may have a substantial effect on the evolution of the orbit, which in turn has an effect on the characteristics of the encounters.
To make sure that the properties of the encounters and the orbit of the Sun in the Galactic potential are consistent, we manually adjust the Sun's position and velocity (and of its OC particles) to its position and velocity on the orbit after each encounter.

\subsection{Captured and lost particles}
\label{sec:captured_and_lost}

After the encounter is simulated, we calculate semi-major axes and eccentricities of the test particles.
A particle is bound to the Sun if its eccentricity with respect to the Sun is $<1$ and its distance from the Sun is smaller than the radius of the Sun's Hill sphere,
$R_{\odot}\left[\mathrm{M}_{\odot}/\left(3M_{\mathrm{G}}(R_{\odot})\right)\right]^{(1/3)}$.
Here, $R_{\odot}$ is the Galactocentric radius of the Sun at the end of the encounter and $M_{\mathrm{G}}(R_{\odot})$ the cumulative mass of the Galaxy within $R_{\odot}$ (Sec.~\ref{sec:gal_pot}).
We calculate the number of particles lost from the Sun's OC 
(initially bound to the Sun, and not bound to it after the encounter), 
and the number of particles captured from the exo-OC of the other star 
(initially bound to the other star, and bound to the Sun after the encounter).
We measure the fractions of OC objects that are lost or captured during an encounter, $N_{\mathrm{loss}}/\noc{}$ and $N_{\mathrm{cap}}/\noc{}$, respectively, where \noc{} is the number of particles in the OC before the encounter.
Finally we remove any particles from the simulation that are $\geq 5$\,pc from the Sun, because they are well outside the Sun's Hill sphere and their dynamics is dominated by the Galactic potential.

\begin{figure*}
\center
\begin{tabular}{c}
\includegraphics[width=0.99\textwidth]{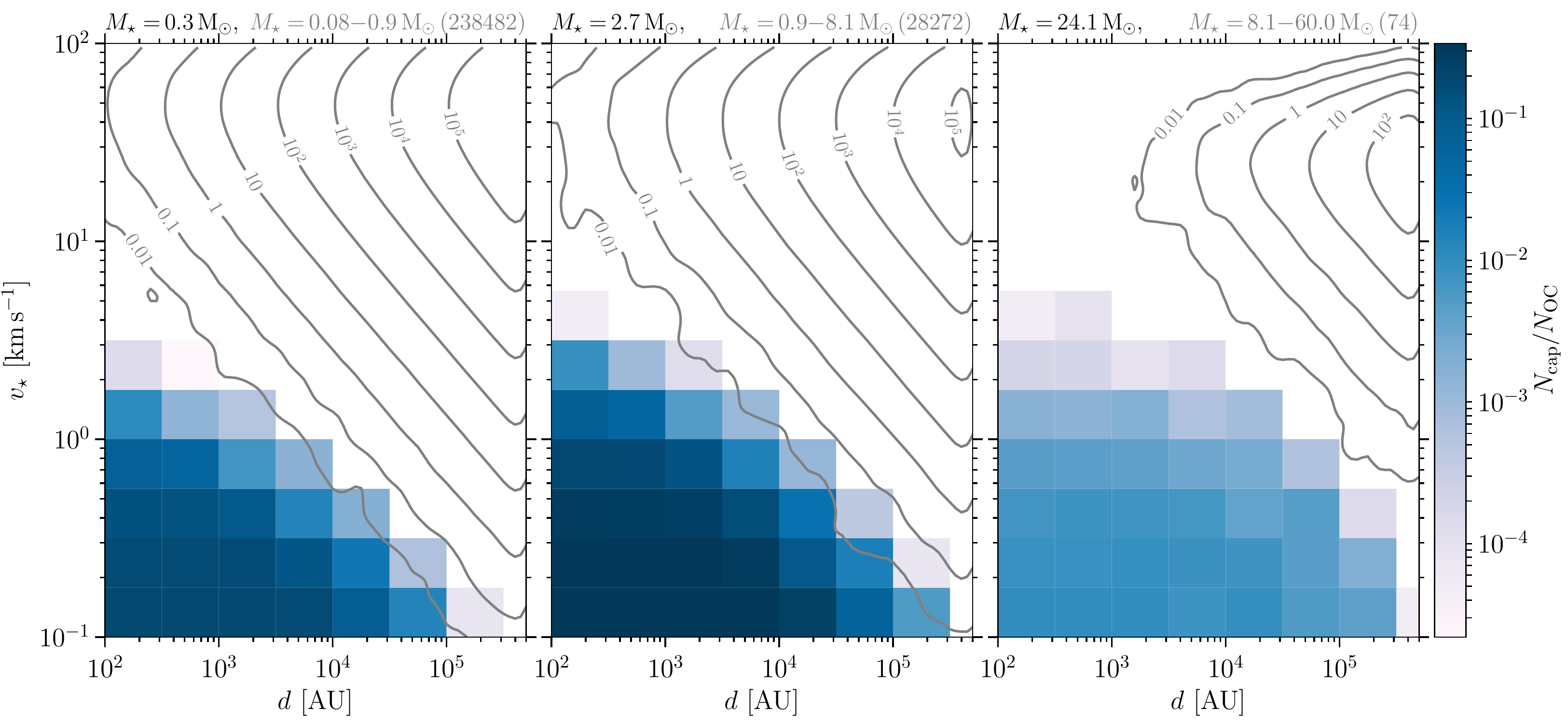} \\
\includegraphics[width=0.99\textwidth]{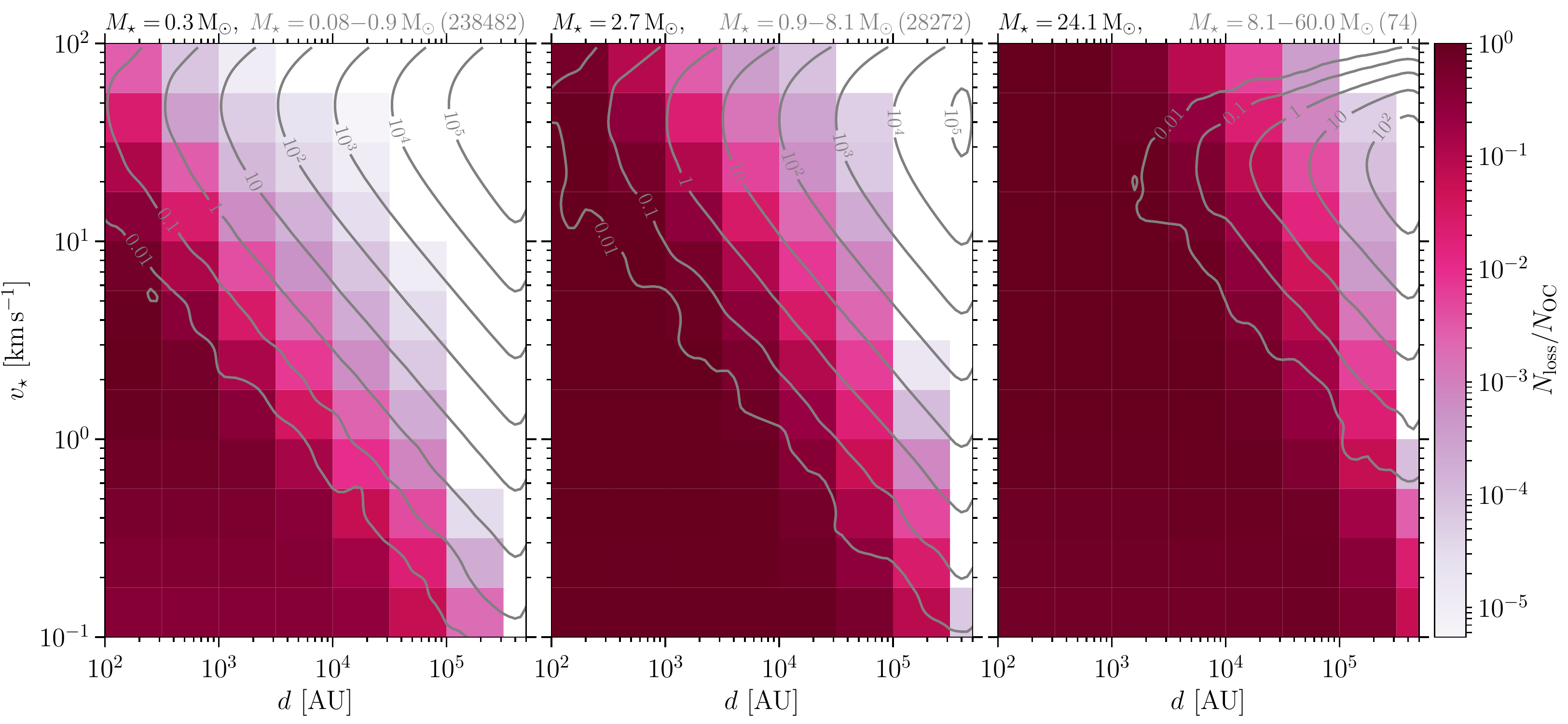}
\end{tabular}
\caption{Fraction of the Sun's OC lost ({\it top}) and captured ({\it bottom}) in encounters with stars of different mass \mstar{}, impact parameter $d$, and velocity $v_{\star}$ is shown by the colour map.
The number density of encounters is shown by the contours.
Plots show results for $\mstar{}=0.3,\,\,2.7,\,\,24.1\,\msun{}$ in {\it left}, {\it middle} and {\it right column}, respectively (\mstar{} is also given above each plot on the left).
The colour scales map $N_{\mathrm{loss}}/\noc{}$ and $N_{\mathrm{cap}}/\noc{}$ for different $d$ and $v_{\star}$.
Note that the colour scale is logarithmic and different for lost and captured fractions.
The contours show the number density of encounters with \mstar{} within the limits given above each plot on the right and with given $d$ and $v_{\star}$ (per $\log (d/\mathrm{AU})\times\log[v_{\star}/(\mathrm{km\,s}^{-1})]$).
The contours are averaged over $10^4$ encounter sets and the average number of stars in each mass range is given in parenthesis above each plot on the right.
}
\label{fig:mdv}
\end{figure*}

\input{table_compare_mdv_and_sets}

\section{Results}
\label{sec:results}

\subsection{Parameter space study of single encounters}
\label{sec:resultsGridsim}

In this section, we describe the results of the parameter space study of individual encounters.
Fig.~\ref{fig:mdv} shows the comets lost from and captured by the Sun's OC ($N_{\mathrm{loss}}/\noc{}$ and $N_{\mathrm{cap}}/\noc{}$) for encounters with stars of different \mstar{}, $d$, and $v_{\star}$, and when the Galactic potential is not taken into account.
We use different numbers of test particles in the clouds for different \mstar{} as specified in Table~\ref{tab:n_mdv}.
The lower limit of $N_{\mathrm{loss}}/\noc{}$ and $N_{\mathrm{cap}}/\noc{}$ is $1/N_{\oc}=0.5$--$5.5\times10^{-6}$ and $1/N_{\mathrm{exoOC}}=0.2$--$2.0\times10^{-6}$, respectively (the exact value depends on \mstar{}).

\subsubsection{Lost and captured fraction of the OC}
\label{sec:lost_captured_oc}

As expected, encounters with lower velocities and smaller impact parameters (when the gravitational interaction is more prolonged and stronger) result in the OC losing and capturing more particles.
Regardless of encounter parameters, a larger fraction of the OC is lost than captured.
There is no capture in encounters with $v_{\star}\gtrsim5$\,\kms{}.
Figure~\ref{fig:mdv} shows that regardless of the encountered star mass, a substantial part of the Sun's OC ($\gtrsim80$\%) is lost in close encounters with $d\sim200$\,AU.
Comparing the three stellar masses $M_{\star}$, the captured fraction is largest for encounters with stars of mass $2.7$\,\msun{} (middle mass bin).
Generally, the binding energy of a comet with a given semi-major axis decreases with decreasing stellar mass.
However, the size of the exo-OC increases with its parent star mass (see scaling given by Eq.~\ref{eq:scaling_ocs}).
More massive stars have bigger exo-OCs with more particles which can be captured at larger distances.

In simulations of individual encounters, the Galactic potential does not substantially affect the captured fractions (see Appendix~\ref{app:mdv_in_potential} for a more detailed description and comparison).
The effect of Galactic tide on the lost fractions of the OC is stronger and depends on the characteristics of the encounter (mostly its velocity).
However, for the majority (more than $85$\%) of encounters simulated in our parameter space study, the difference between $N_{\mathrm{loss}}/\noc{}$ in the simulations with and without the effect of the Galactic tide is less than $5$\%.
This is expected, because the gravitational dynamics within the OC's outer radius of $10^5$\,AU (Sec.~\ref{sec:oc}) is dominated by the Sun and the stars it encounters. 
The timescale on which the Galactic tide affects the cometary orbits (causing oscillations of eccentricity and inclination with period on the order of $\sim1$\,Gyr, e.g. \citealp{1986Icar...65...13H}, \citealp{2010A&A...516A..72B}) is also much longer than the encounter timescale.

\subsubsection{Comparison with encounter sets}
\label{sec:comp_enc_set}

By comparing the captured and lost fractions in the parameter space study with the encounter sets, we measure the average captured and lost fractions of the OC over the Solar orbit.
Table~\ref{tab:n_enc_per_orbit} shows the comparison.
The results are averaged over $1000$ different encounter sets for each type of orbit.
We count the number of encounters in the bins that are used in the parameter space study (Fig.~\ref{fig:mdv}).
Out of $3000$ sets there are only 102 encounters that are within the parameter space study bins for which the Sun captures particles from the passing star.
There is always only one such encounter per set. 
In Fig.~\ref{fig:mdv}, these encounters come from the regions with $v_{\star}$ and $d$ where the $0.1$ and $0.01$ contours of the normalized number of encounters overlap with $N_{\mathrm{cap}}>0$.
It turns out to be efficient to capture exocomets during flybys with lower-mass stars.
For each type of orbit, we calculate the average captured fraction of the OC, $\eta_{\mathrm{cap}}$, as the average $N_{\mathrm{cap}}/\noc{}$ over the $1000$ encounter sets.
The values $\eta_{\mathrm{cap}}$ are about $6\times10^{-5}$, $1.4\times10^{-4}$, and $1.6\times10^{-4}$ for the orbit migrating inwards, not migrating, and migrating outwards.
Note that these captured fractions are derived from simulations of single encounters and do not take the subsequent evolution of the OC into account.
The captured exocomets are often again lost from the OC some time after the encounter that delivered them.
We discuss the evolution of the captured exocomets during the Solar orbit in Sec.~\ref{sec:resultsOrbitsim}.

In Fig.~\ref{fig:mdv} we show that for all three masses of the encountered star, a substantial part of the Sun's OC ($\gtrsim80$\%) is lost in a close encounter with $d\sim200$\,AU.
Regardless of the migration of the Solar orbit, the highest fraction of the OC is lost during flybys with stars of $\mstar{}=0.9$--$8.1$\,\msun{}.
Encounters with low-mass stars ($\mstar{}<0.9$\,\msun{}) on the other hand, result in a loss of $\lesssim3$\% of the OC.
The average fraction of the OC lost over the Sun's lifetime, $\eta_{\mathrm{loss}}$, is $32$\%, $46$\%, and $63$\% for the orbit migrating inwards, not migrating, and migrating outwards.

The averaged fractions captured or lost from the OC differ for different migration types of the Solar orbit.
However, the rates of encounters that result in capturing exocomets ($N_{\mathrm{enc}}^{\mathrm{cap}}/N_{\mathrm{enc}}$ in Table~\ref{tab:n_enc_per_orbit}) for given mass of the passing star are similar and do not depend on the migration of the Solar orbit.

\subsection{Full orbit simulations}
\label{sec:resultsOrbitsim}

\begin{figure}
\center
\includegraphics[width=0.49\textwidth]{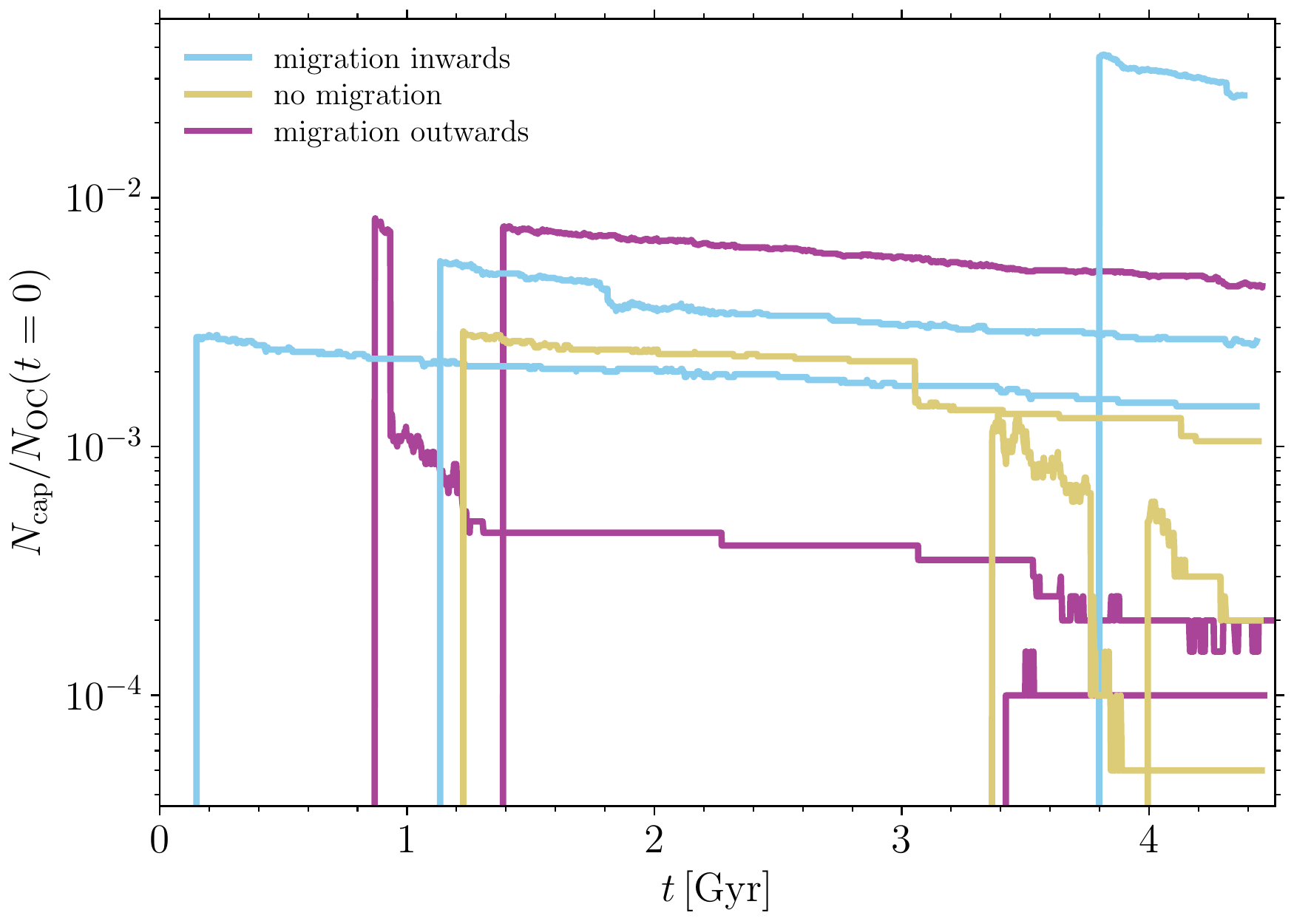}
\caption{Evolution of the captured fraction of the OC.
Note that here $N_{\mathrm{OC}}(t=0)=2\times10^4$ is the number of OC's comets at the beginning of the simulation.
Three encounter series are shown for each type of orbit.
}
\label{fig:gaint_orbits}
\end{figure}

Using the results of the parameter space study we pick 30 encounter series that each contains an encounter we expect result in exocomet capture (because its parameters fall within a bin of the parameter space study that has $N_{\mathrm{cap}}>0$).
We simulate the orbit of the Sun and its OC under the influence of the encounters as described in Sec.~\ref{sec:orbitsim}.
Out of those, 17 encounter series result in $N_{\mathrm{cap}}>0$.
For all these 17 encounter series, the exocomets are captured during the encounter that we identified using the parameter space study.
For the remaining (13) series, exocomets are unbound from their parent star but are outside the Sun's Hill sphere after the encounter.
The number of exocomets in the OC decreases after their acquisition due to stellar flybys and the effect of the Galactic tide.
Only for 12 orbits are some exocomets retained in the Sun's OC until the end of the simulation.
The other five orbits include encounters that result in the OC capturing exocomets that are later lost.
How long exocomets are retained in the OC depends on their number, orbits around the Sun, and the parameters of the encounters that follow.
For one of those orbits, the OC retains the captured exocomet for three encounters only (corresponding to $\sim0.9$\,Myr), while exocomets captured in other orbit are lost only after $450$ encounters ($\sim0.7$\,Gyr).
In some cases a substantial number of exocomets is lost at once due to a strong encounter.

We run six additional simulations.
For each type of orbit, we simulate a random encounter set that does not result in exocomet capture. 
Finally, we simulate a solar orbit with the OC not influenced by any encounters (that is, the change in $N_{\mathrm{OC}}$ is only due to the effect of the Galactic tide) for each type of orbit.

In Fig.~\ref{fig:gaint_orbits} we show the evolution of OC's captured fraction for nine encounter series (three per type of orbit) that retain a finite number of exocomets until the current time ($t=4.5\,$Gyr).
There is no obvious difference in the  captured fraction evolution for different types of orbit.
Note that here we measure the fraction of the OC at $t=0$ of the complete orbit simulation, that is $N_{\mathrm{OC}}(t=0)=2\times10^4$.
For all of the nine orbits, the captured exocomets constitute only a small fraction of the initial OC: typically about $0.2$\% and $3$\% at maximum.
If the number of captured exocomets is compared with the total number of comets at the moment of the encounter, the fractions are slightly higher because the OC has already lost some of its comets.
This fraction is typically about 0.5\%, 10\% at maximum, and depends on the time the encounter occurred as well as the effect of the earlier encounters.

\begin{figure}
\center
\includegraphics[width=0.49\textwidth]{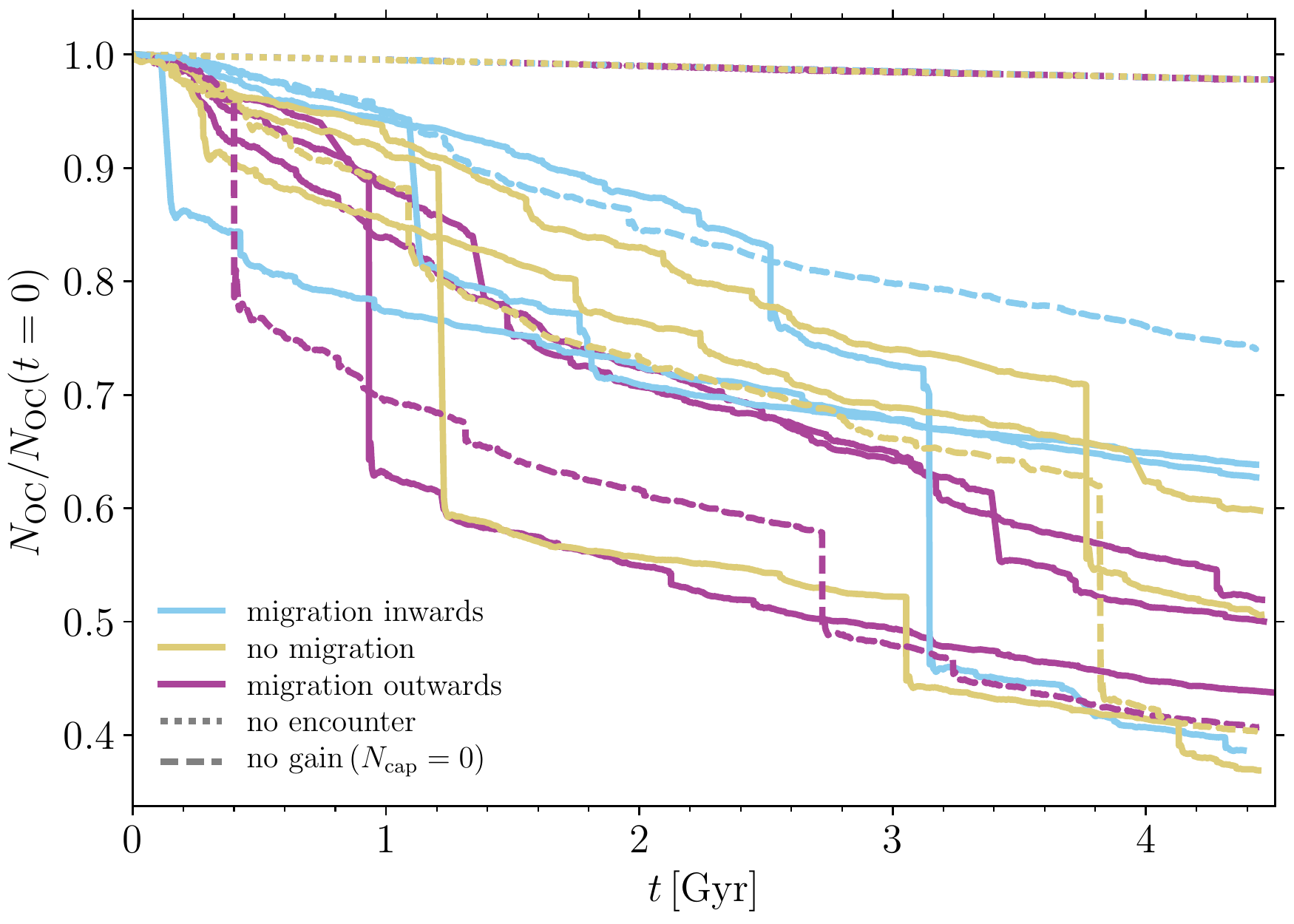}
\caption{Evolution of the number of particles (comets and exocomets) in the OC.
$N_{\mathrm{OC}}(t=0)$ is measured at the beginning of the simulation here.
The same nine orbits with exocomets as in Fig.~\ref{fig:gaint_orbits} are shown by full lines.
The dashed lines show examples of orbits during which no exocomets were transferred to the Sun's OC.
The dotted lines show the evolution of an orbit without encounters where the OC evolves only under influence of the Galactic tide (note that a line is shown for each orbit type but they are virtually indistinguishable).
}
\label{fig:nt_orbits}
\end{figure}

In Fig.~\ref{fig:nt_orbits} we show the evolution of the total number of particles (comets and exocomets) in the OC for the same nine orbits as in Fig~\ref{fig:gaint_orbits}.
It is important to keep in mind that those nine orbits are picked manually such that their encounter series include an exocomet-capturing encounter.
This flyby has generally low velocity and relatively small separation and also results in a substantial loss of comets from the OC.
We therefore also plot three additional orbits with no captured exocomets, for which the encounter sets are random (dashed lines).
Except for the orbit migrating inwards, these orbits also experience at least one encounter that results in a loss of about $20$\% of the OC.
The OC loses between about $25$\% and $65$\% of its mass for each orbit in Fig.~\ref{fig:nt_orbits}.
We also plot orbits with the OC not influenced by any encounters for each type of orbit.
In these cases the OC loses about $2$\% of its comets.

More simulations would be needed to judge if the total fraction of the OC mass lost depends on radial migration.
Comparison of the line slopes for different orbit types in Fig.~\ref{fig:nt_orbits} indicates that unless perturbed by a strong encounter, $N_{\mathrm{OC}}$ decreases slightly faster for orbits starting at smaller Galactocentric radii.
However, for the majority of orbits in Fig.~\ref{fig:nt_orbits}, the evolution of the lost fraction is more influenced by a few strong encounters than by radial migration.

\subsection{Orbital elements of lost comets and captured exocomets}
\label{sec:exo_oc_elements}

\begin{figure*}
\center
\includegraphics[width=0.33\textwidth]{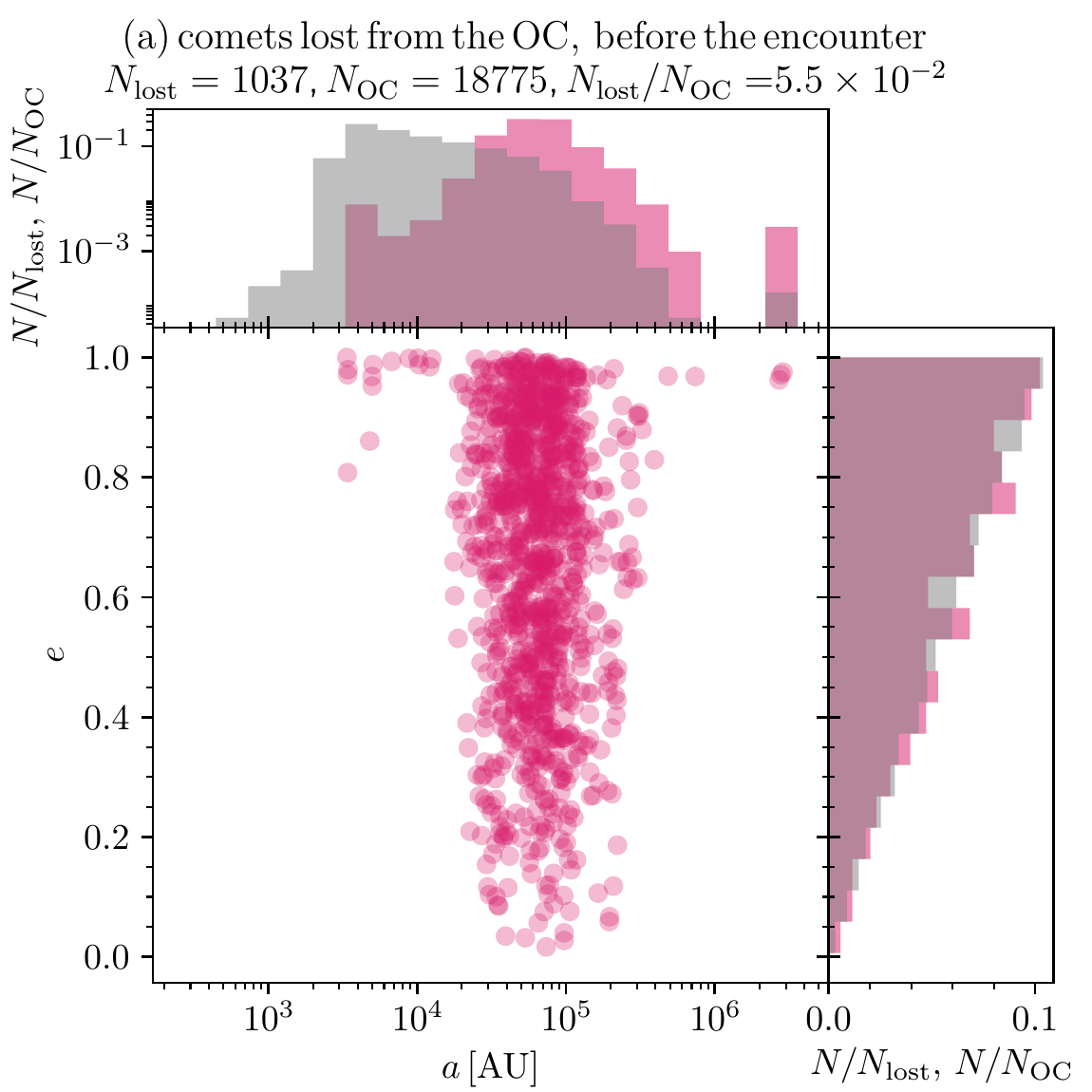} \hspace{-0.2cm}
\includegraphics[width=0.33\textwidth]{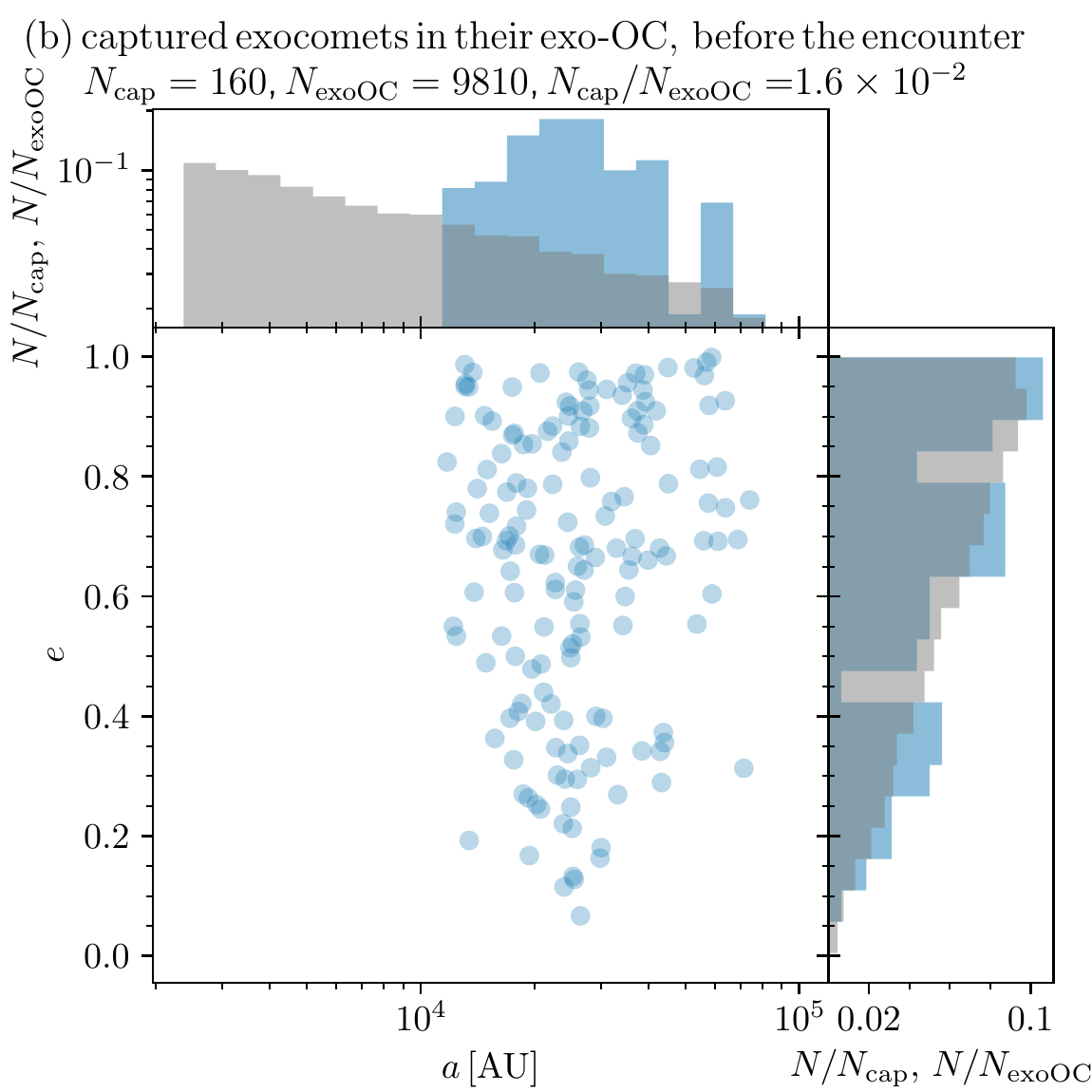} \hspace{-0.2cm}
\includegraphics[width=0.33\textwidth]{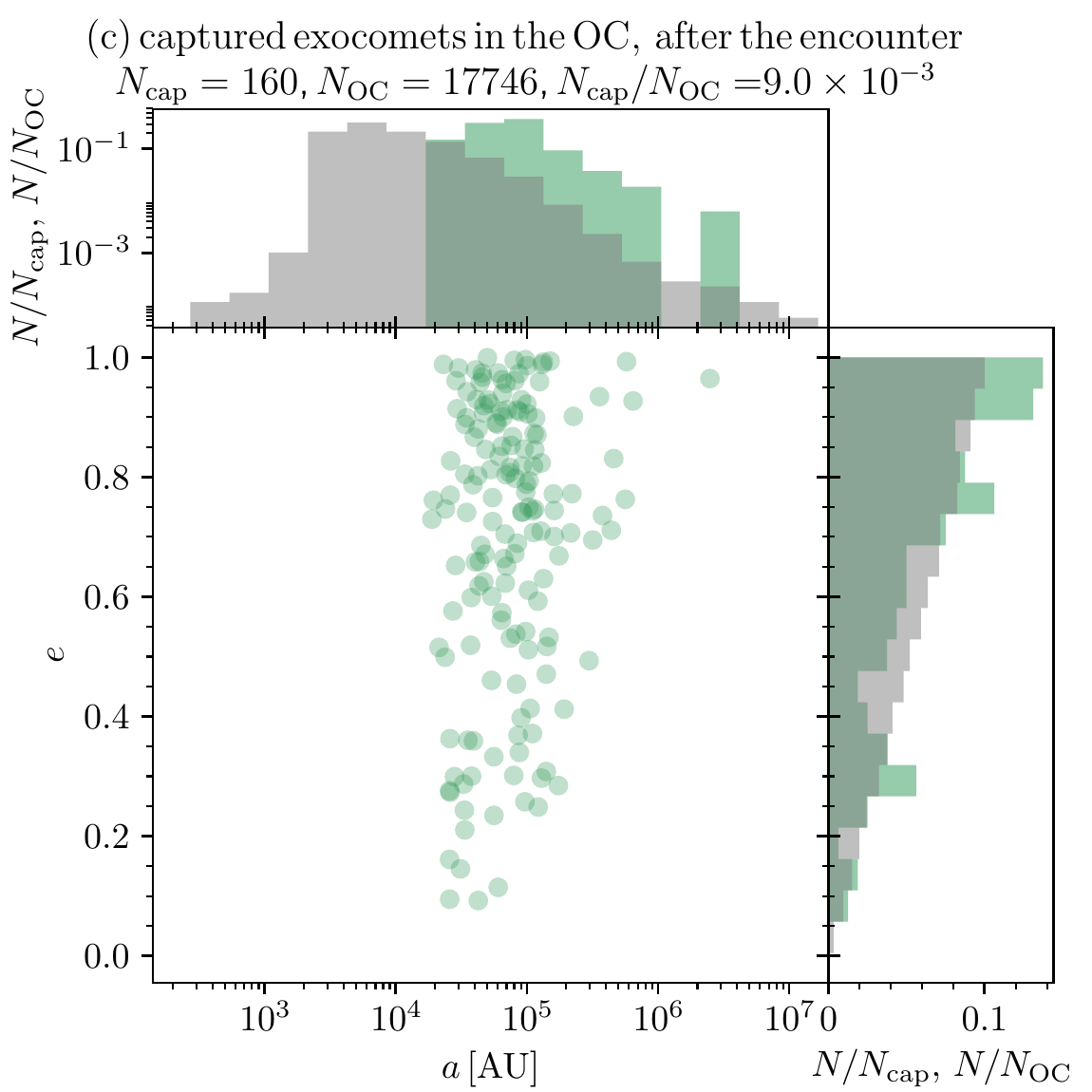}
\caption{Semi-major axes and eccentricities of (exo)comets that were lost and captured during an encounter with $M_{\star}=0.5\,\msun{}$, $d=7.7\times10^4$\,AU, and $v_{\star}=0.18$\,\kms{} at $t=0.9$\,Gyr.
Some (four) of the captured exocomets are retained in the OC until the end of the simulation.
Plot (a) shows the distribution of lost comets in the OC before the encounter.
The lost comets are depicted by the pink points and histograms while the grey histograms are for all comets in the OC.
Plot (b) shows the distribution of the captured exocomets in their parent exo-OC before the encounter.
The captured exocomets are depicted by the blue points and histograms while the grey histograms are for all exocomets in the exo-OC.
Plot (c) shows the distribution of the captured exocomets in the OC after the encounter.
The exocomets are depicted by the green points and histograms while the grey histograms are for all comets in the OC.
Here \noc{} includes both the comets native to the OC and the captured exocomets.
All the histograms are normalized by the number of particles, but not by the bin size.
The numbers of lost and captured (exo)comets are indicated above each plot.
Note that the OC also captures some of its own comets that were stripped before the encounter and live just outside the potential well of the Sun.
}
\label{fig:ae_surviving}
\end{figure*}

In Fig.~\ref{fig:ae_surviving}, we show the semi-major axes and eccentricities of comets lost and exocomets captured during an example encounter taken from a full orbit simulation.

Some of the exocomets captured in Fig.~\ref{fig:ae_surviving} are retained in the OC until the end of the simulation. 
The field star has $M_{\star}=0.5\,\msun{}$, $d=7.7\times10^4$\,AU, and $v_{\star}=0.18\,\kms{}$ and the encounter occurs at $t=0.9$\,Gyr of an outward migrating orbit.
This is an encounter with a low-mass star at a small separation and low velocity which is relatively rare (see contours in Fig.~\ref{fig:mdv}). 
The orbit that includes the encounter is one of those plotted in Fig.~\ref{fig:gaint_orbits} (see one of the violet lines with suddenly increasing $N_{\mathrm{cap}}$ at $\sim0.9$\,Gyr) and Fig.~\ref{fig:nt_orbits}.
Fig.~\ref{fig:ae_surviving}a shows that during this encounter $5.5$\% of comets are lost from the OC.
The distribution of the semi-major axis of lost comets (pink histograms) is concentrated at larger semi-major axes ($a\sim8\times10^4$\,AU) than considering all comets in the OC (grey histograms).
The distribution of lost comets eccentricities is very similar to the one for all comets.
From Fig.~\ref{fig:ae_surviving}b we see that exocomets are captured from the outer part of their parent exo-OC.
After their capture to the OC (Fig.~\ref{fig:ae_surviving}c), the majority of exocomets have orbits with $a=2\times10^4$--$10^5$\,AU and are on average slightly more eccentric than the OC comets.

The distributions we show in Fig.~\ref{fig:ae_surviving} are only an example of the possible encounter outcome.
Fig.~\ref{fig:ae_surviving} seems to indicate that the captured exocomets and lost comets come from a specific parts of the clouds, but in general they will come from different parts of the clouds.
Similarly, the exocometary orbits in the OC can have various semi-major axes and eccentricities.

\subsection{Error estimates}
\label{sec:errors}

Here we discuss errors resulting from numerical limitations and our parameter choices.

We carry out additional simulations of individual encounters that result in the OC capturing exocomets using 10-times more particles than in the standard runs of our parameter space study (Table~\ref{tab:n_mdv}).
We draw 200 random sub-samples with a tenth of the number of particles each and we measure the standard deviations of their $N_{\mathrm{cap}}/\noc{}$ and $N_{\mathrm{loss}}/\noc{}$.
From those we estimate the uncertainty of $N_{\mathrm{cap}}/\noc{}$ and $N_{\mathrm{loss}}/\noc{}$ to be $20$ and $5$\%, respectively.

To couple the Galactic tide to the system of the Sun, the field star and their clouds, we set the \textsc{bridge} timestep to $1$\,Myr (see Sec.~\ref{sec:gal_pot}).
To estimate the error this choice introduces, we run several simulations using shorter timesteps.
First, we run simulations of the non-migrating orbit without the effect of stellar encounters using \textsc{bridge} timesteps of $1$, $0.5$, $0.1$, $5\times10^{-2}$, and $1\times10^{-2}$\,Myr and $N_{\mathrm{OC}}=5\times10^3$.
After $t=2$\,Gyr the fraction of comets retained in the OC for the three latter timesteps differs by $\lesssim0.5$\%.
Results for the $1$\,Myr timestep differ from the shortest timestep by $\sim1$\%.
Further, we run simulations of the Solar orbit (all three types, with $N_{\mathrm{OC}}=10^4$) in the Galactic potential without the effect of stellar encounters.
We follow the number of comets in the OC in time and compare OC's erosion due to the Galactic tide.
At the end of the simulations ($t=4.5$\,Gyr), the difference between the fraction of comets retained in the OC for the timestep of $1$\,Myr and $0.1$\,Myr is $\lesssim 2$\%.
Finally we run one full orbit simulation (including the passing stars and their clouds) with the timestep of $0.1$\,Myr.
The difference in the captured fraction of the OC $N_{\mathrm{cap}}/\noc{}$ is $\lesssim3\times10^{-3}$.
This however, corresponds to up to $1.6$-times more captured exocomets for the longer timestep ($91$ and $152$ for $0.1$ and $1$\,Myr, respectively).

\section{Discussion}

The Sun's OC is not an isolated system. 
In section~\ref{sec:results} we have investigated the loss of comets from interactions with the Galactic tidal field and stellar encounters, and the transfer of exocomets form exo-OC hypothesized around passing field stars.
We find the transfer of exocomets is rare and the main reason for this lies in the paucity of low velocity encounters.
This is in contrast to the loss of comets from the OC, for which stars passing by with higher velocity are also efficient.
Still, this loss of comets is mainly mediated (sec.~\ref{sec:resultsOrbitsim}) by strongly interacting encounters.  This implicitly raises a number of issues with respect to our simulation results: 
(1) how good was the selection we made in section~\ref{sec:enc_selection}) of relevant encounters (and can we improve upon this),
(2) how robust are our results with respect to numerical parameter choices, and 
(3) how strongly do our results depend on the assumptions pertaining to the stellar velocity distributions? Below we will examine these questions in more detail.

\subsection{Proxies for the lost or captured fractions of the OC}
\label{sec:proxies}

In Sec.~\ref{sec:enc_selection}, we describe the selection of the simulated encounters using function $M_{\star}/(v_{\star}d)$, which is proportional to the impulse gained by the Sun during an encounter.
While this selection seems reasonable, it does raise the question whether a different selection function would yield a more accurate prediction for the fraction of the Sun's OC lost and captured. 
To examine this we investigate proxies of the form
\begin{equation} 
\label{eq:proxy}
\begin{array}{l}
f(M_{\star},v_{\star},d)=\alpha \mstar{}^{a}d^{b}v_{\star}^{c}, \\
N_{X}/\noc{}=
\begin{cases}
    f(M_{\star},v_{\star},d)    & \quad \text{if } f(M_{\star},v_{\star},d) < 1, \\
    1                           & \quad \text{if } f(M_{\star},v_{\star},d) \geq 1. \\
\end{cases}
\end{array}
\end{equation}
Here $N_{X}$ represents the fraction of OC that is lost ($N_{\mathrm{loss}}$) or captured ($N_{\mathrm{cap}}$) during an encounter characterized by $M_{\star}$ in Solar masses, $d$ in AU, and $v_{\star}$ in $\kms{}$.
The number of particles in the OC before the encounter is \noc{}, and $a$, $b$, and $c$ are free parameters and $\alpha$ is a scaling parameter.

We fit these parameters in Sec.~\ref{sec:proxies_fit}, for both lost and captured OC fractions, using the results of the parameter space study (Sec.~\ref{sec:lost_captured_oc}).
In Sec.~\ref{sec:resultsOrbitsim_proxy}, we apply the proxies to the results of the full orbit simulations (presented in Sec.~\ref{sec:resultsOrbitsim}).

\subsubsection{Fitting proxies using the parameter space study}
\label{sec:proxies_fit}

\begin{figure}
\center
\includegraphics[width=0.35\textwidth]{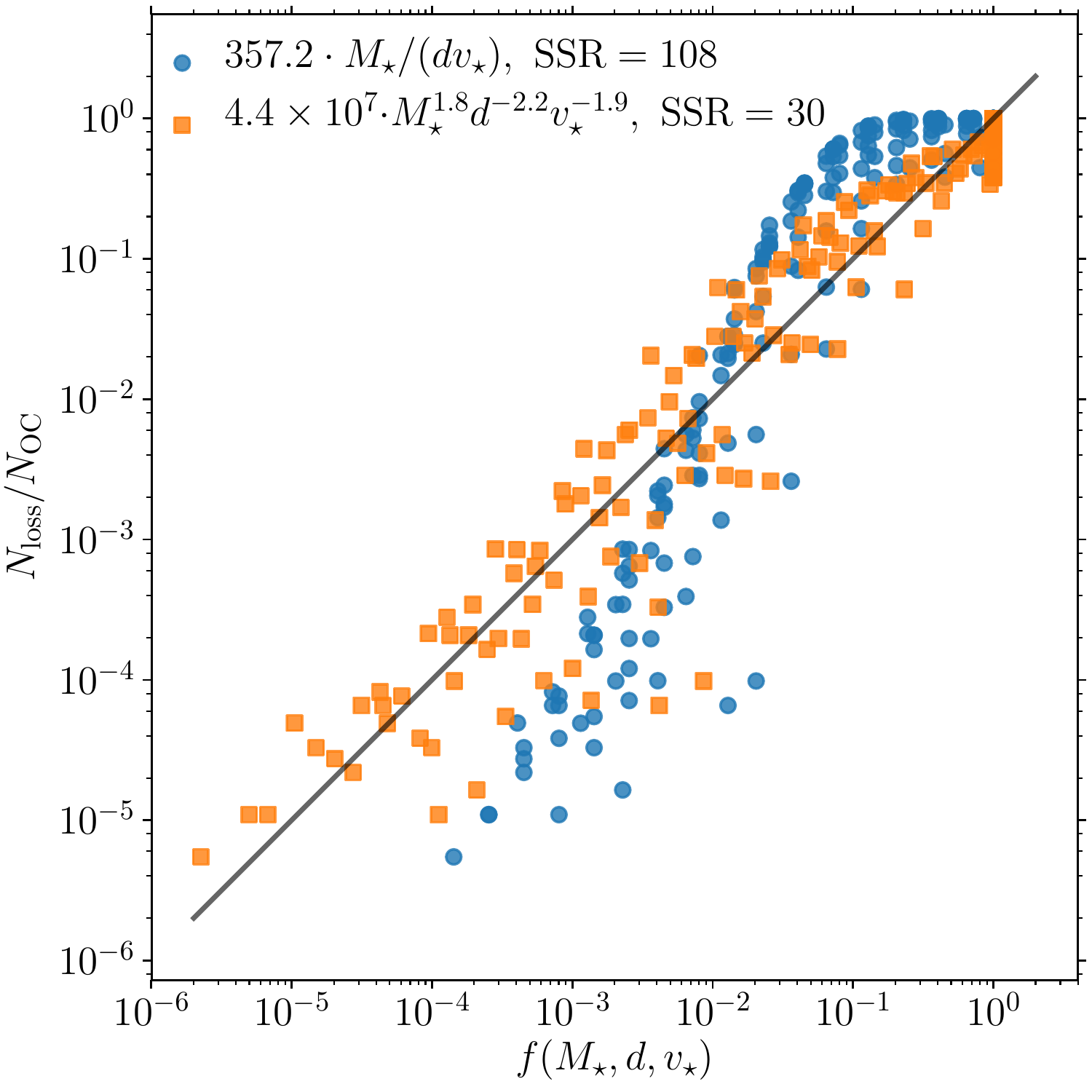} \\
\includegraphics[width=0.35\textwidth]{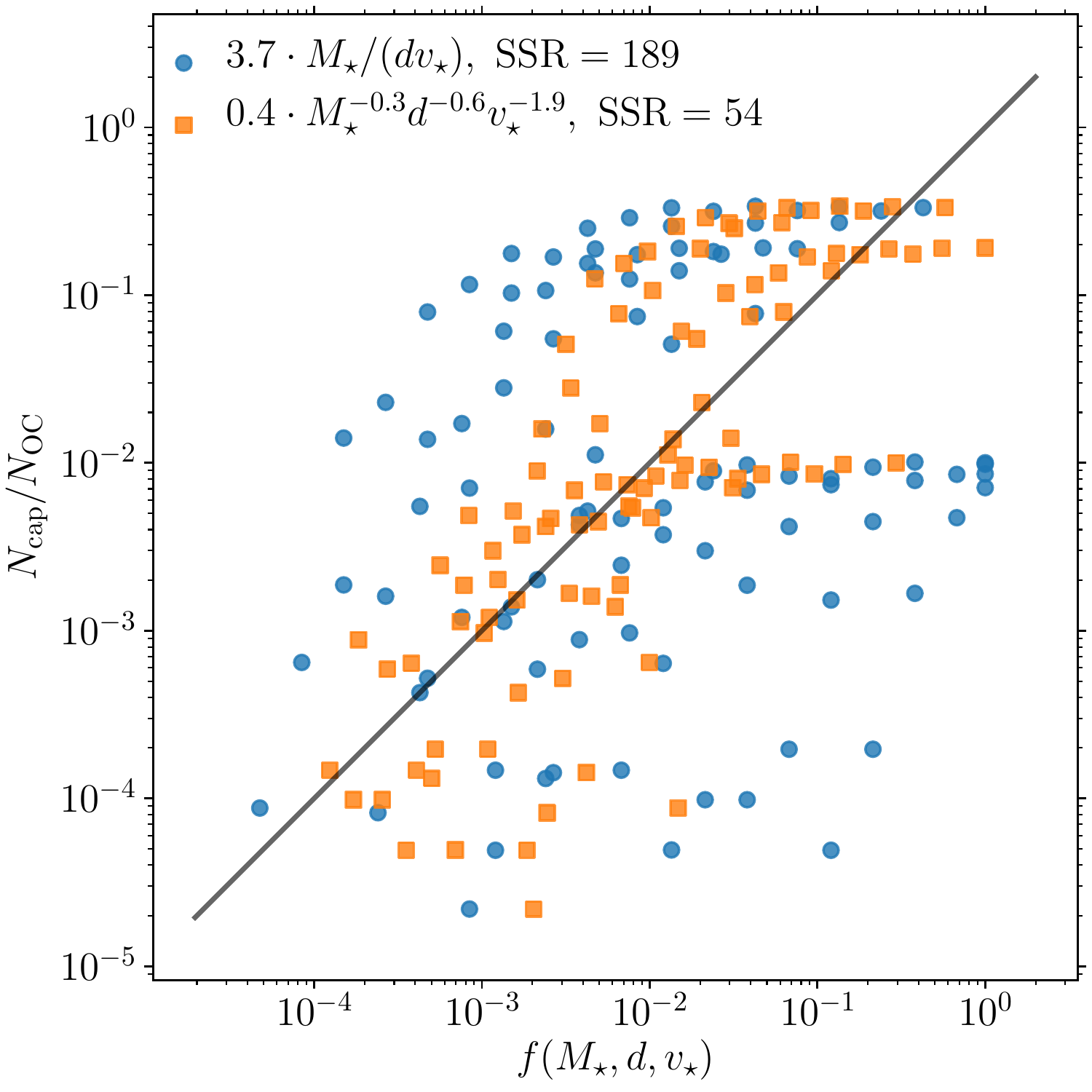}
\caption{Proxies for the lost ({\it top}) and captured ({\it bottom}) fractions of the OC.
The values for the lost and captured fractions from the parameter space study are on the vertical axis.
The values for the proxies are on the horizontal axis.
The blue circles show a function linearly proportional to the impulse gained by the Sun, $\propto\mstar{}/(dv_{\star})$.
Orange squares show a more general function of a form given by Eq.~\ref{eq:proxy} using the optimal coefficients (Table~\ref{tab:proxies}).
The full line is one-to-one correlation.
Both axes are logarithmic.
}
\label{fig:proxies}
\end{figure}

Because the fraction of OC objects lost and captured range over several orders of magnitudes, we perform fitting in log--log space.
We fit the logarithm of the function given by Eq.~\ref{eq:proxy} (i.e. $\log(N_{X}/\noc{})$) to the logarithm of the simulated results, $\log(N_{\mathrm{loss}}/\noc{})$ and $\log(N_{\mathrm{cap}}/\noc{})$.
We do not consider simulations in which no particles are lost or captured. 
The optimal values and their uncertainties as estimated by a non-linear least squares method are given in Table~\ref{tab:proxies}.
In Fig.~\ref{fig:proxies}, we show the comparison between the fitted proxies and the function $\beta M_{\star}/(v_{\star}d)$, for which we fitted the scaling constant $\beta$.
The sum of squared residuals (SSR) of the fits for the proxy and for $\beta M_{\star}/(v_{\star}d)$ are $37$ and $115$ in case of the lost fraction, and $46$ and $185$ for the captured fraction (residuals are also calculated in log--log space).
This indicates that the proxies given by Eq.~\ref{eq:proxy} are a better prediction for the lost and captured fractions of the OC than a function $\propto M_{\star}/(v_{\star}d)$.

Note that the fitted values of the parameters for the lost comets are close to $(M_{\star}^2/(v_{\star}^2 d^2)$. 
This implies that the dependence is close to the impulse squared rather than the impulse. 
Furthermore notice that the scatter in the relation for the captured exocomets is much bigger than for the lost comets (Fig.~\ref{fig:proxies}). 
This could be due to another parameter independent from $M_{\star}$, $d$, and $v_{\star}$ being important in determining the capture rate (for example orbital parameters of the exocomets) or due to the much less captured than lost particles.

\begin{table}
\caption{Optimal values of the free parameters of the proxy given by Eq.~\ref{eq:proxy} for the lost and captured fraction of the OC during a stellar encounter.}\label{tab:proxies}
\center
\begin{tabular}{r*{4}{@{\hspace*{0.2cm}}r@{$\pm$}l}}
\hline
        & \multicolumn{2}{c}{$a$} 
        & \multicolumn{2}{c}{$b$} 
        & \multicolumn{2}{c}{$c$} 
        & \multicolumn{2}{c}{$\alpha$} \\
lost    & $1.84$&$0.05$   & $-2.17$&$0.06$    & $-1.91$&$0.05$    & $(4.4$&$2.5)\times10^7$   \\
captured    & $-0.3$&$0.1$    & $-0.63$&$0.09$    & $-1.9$&$0.2$      & $0.4$&$0.3$   \\
\hline
\end{tabular}
\end{table}

\subsubsection{Testing proxies on the full orbit simulations}
\label{sec:resultsOrbitsim_proxy}

\begin{figure}
\center
\includegraphics[width=0.49\textwidth]{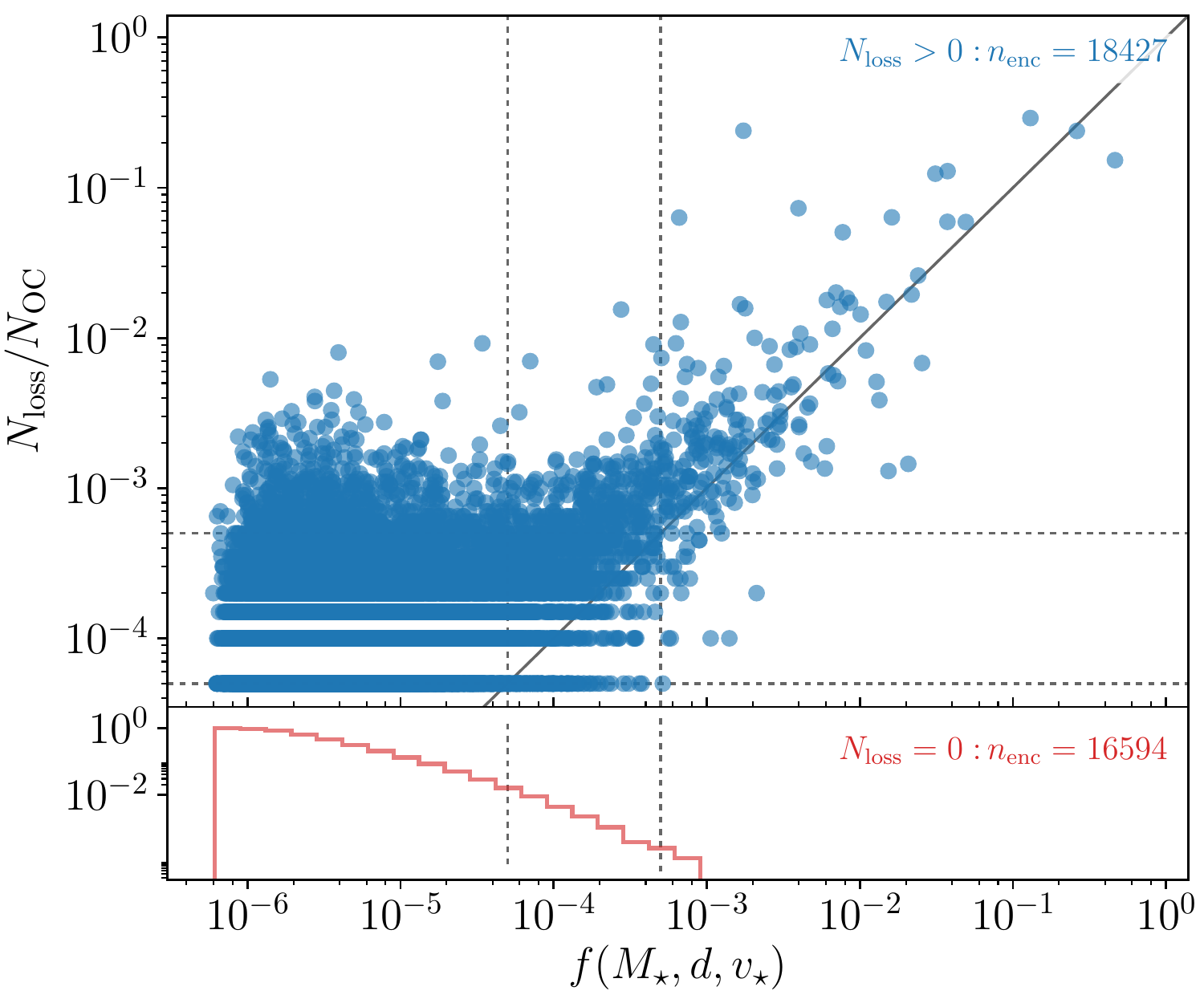}
\caption{Fraction of the OC lost during the nine simulated encounter series presented in Figs.~\ref{fig:gaint_orbits} and \ref{fig:nt_orbits}.
The plot has two panels.
In both, the horizontal axis shows the proxies (Eq.~\ref{eq:proxy} with the coefficients from Table~\ref{tab:proxies}).
The top panel shows encounter simulations that result in OC losing particles ($N_{\mathrm{loss}}>0$).
The bottom panel shows cumulative probability density function of the encounters without any loss. 
The number of plotted encounters is indicated in the upper right corner of each panel.
The dashed lines indicate corresponding to one (i.e. the lower limit of $5\times10^{-5}$) and ten particles lost from the OC.
The full line is one-to-one correlation.
}
\label{fig:proxy_orbits_loss}
\end{figure}

\begin{figure}
\center
\includegraphics[width=0.35\textwidth]{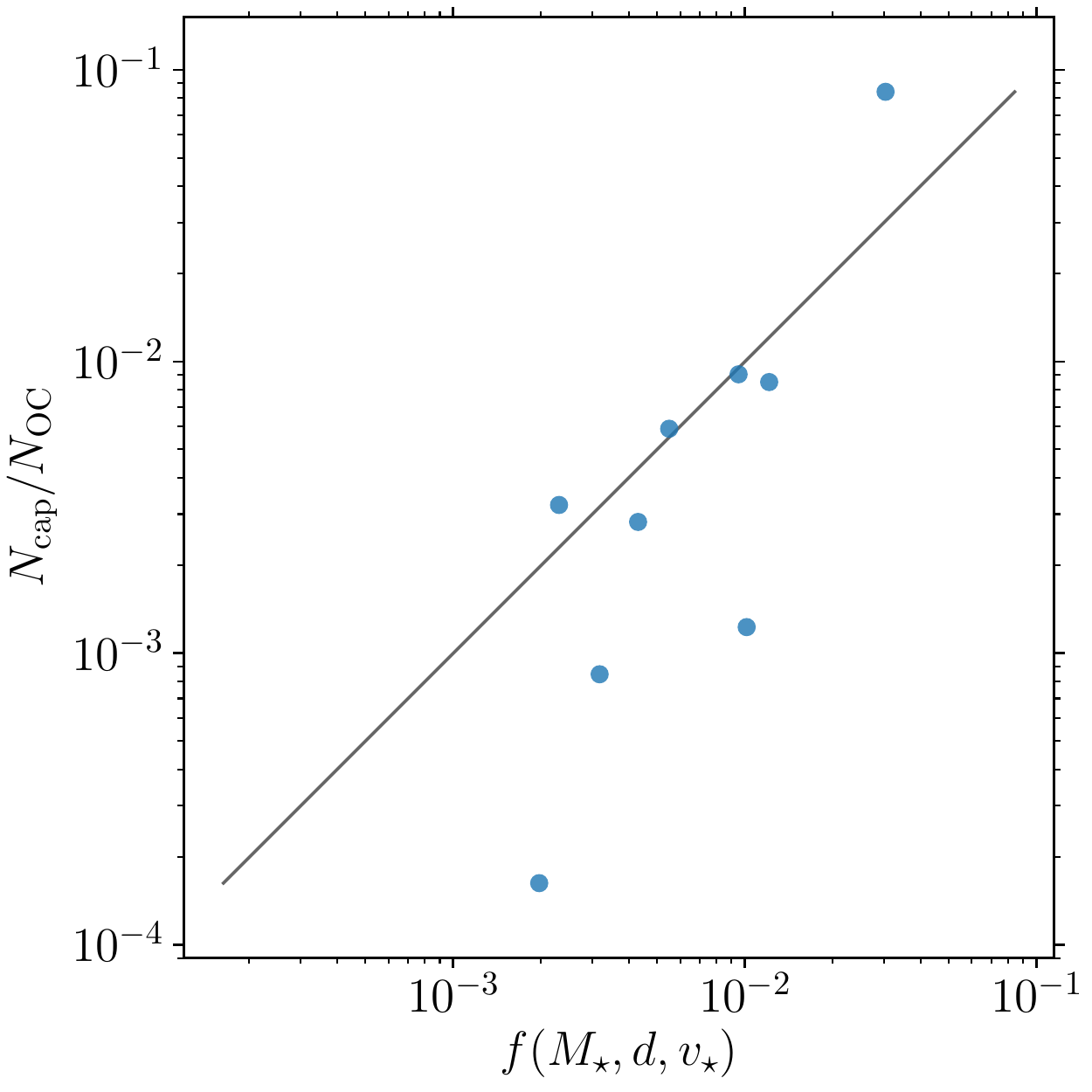}
\caption{Fraction of the OC captured during the nine simulated encounter series presented in Figs.~\ref{fig:gaint_orbits} and \ref{fig:nt_orbits}.
The horizontal axis shows the proxies (Eq.~\ref{eq:proxy} with the coefficients from Table~\ref{tab:proxies}) while the vertical axis shows the simulated captured fractions.
Note that here \noc{} corresponds to the number of comets in the OC directly before the encounter (unlike in Figs.~\ref{fig:gaint_orbits} where the initial number of OC comets, $\noc{}(t=0)$, is used).
The full line is one-to-one correlation.
}
\label{fig:proxy_orbits_gain}
\end{figure}

We use Eq.~\ref{eq:proxy} with the coefficients in Table~\ref{tab:proxies} to calculate proxies for the captured and lost fractions of the OC for the nine simulated encounter series presented in Figs.~\ref{fig:gaint_orbits} and \ref{fig:nt_orbits}.
In Figs.~\ref{fig:proxy_orbits_loss} and \ref{fig:proxy_orbits_gain}, we compare the proxies and the simulated lost and captured fractions of the OC, respectively.

The proxies for the lost fraction show substantial scatter and except for encounters resulting in $N_{\mathrm{loss}}/\noc{}\gtrsim10^{-2}$ typically underestimate the loss.
This is probably due to the Galactic tide, which is not taken into account when fitting the proxy function, and which results in an additional loss of comets that are only weakly bound after the encounter (see Appendix~\ref{app:mdv_in_potential}).
In the bottom panel of Fig.~\ref{fig:proxy_orbits_loss}, we show the cumulative probability density distribution of the proxies for the encounters that do not result in any loss of comets.
There are very few cases ($\lesssim1$\%) for which the proxy is higher than the simulation resolution limit ($5\times10^{-5}$) and all the proxies $\lesssim10^{-3}$.
In Fig.~\ref{fig:proxy_orbits_gain} we show that the proxy derived in Sec.~\ref{sec:proxies} gives a good prediction for the captured fraction of the OC.

\subsection{Model assumptions}
\label{sec:model_assumtions}

A number of (explicit or implicit) model assumptions enter in our model.

First, we assume that every star is surrounded by an exo-OC. 
This assumption results from the Copernican principle, in this case implying that the Solar system's OC is not special. 
In reality this may well overestimate exo-OCs masses.
The production of a (exo)cometary cloud may require a specific configuration of the planetary system and its surrounding environment and might be rather rare \citep{2017MNRAS.464.3385W}.
From this point of view, our results on exocometary transfer represent an upper limit. 
On the other hand, if exo-OCs are a rare occurrence, even the relatively modest transfer rates we find may represent a significant seeding to exo-OC-poor stars from stars with exo-OC of a high mass.

Second, our assumption that only one star at a time is interacting with the Sun may affect the quantitative results for slow encounters (which take
longest and are also most important for captures). 
It may be that during such a slow encounter, comets that are unbound from the low velocity star due to another star passing by with higher velocity are captured.
Such multiple interaction could possibly also allow for extra captures from wide encounters. 
In any case, our findings with respect the the cometary losses are robust and point to a considerable erosion of the OC from stellar encounters. 

Summarizing, we argue that our encounter series include all the important encounters that result in a substantial fraction of the OC being lost or captured.
We select simulated encounters based on the impulse they deliver to the Sun (Sec.~\ref{sec:enc_selection}).
\citet{1976BAICz..27...92R} showed that when considered over the Solar lifetime, the total impulse delivered to the Sun is dominated by the few closest encounters (with pericenter $<2\times10^4\,$AU).
\citet[][Sec.~2.3]{2011Icar..215..491K} 
confirmed that the formation of the OC (mainly the location of its inner edge) is substantially affected by a handful of strong isolated stellar encounters.
These findings indicate that our selection of simulated encounters (about $3500$--$4000$) includes those that substantially affect the Sun's OC.
Further, the impulse gained by the Sun in an encounter compares reasonably well with the proxies for lost and captured fractions of the OC (Sec.~\ref{sec:proxies}, Fig.~\ref{fig:proxies}).
Finally, only about half of the simulated encounters result in the OC losing particles (see Fig.~\ref{fig:proxy_orbits_loss}).

\subsection{Planetary perturbations and the longterm evolution of the OC}
\label{sec:planetary_perturb}

Planetary perturbations play an important role in the production of the observable comets (\citealp{1950BAN....11...91O}, \citealp{2009Sci...325.1234K}, or \citealp{2014Icar..231..110F}) as well as in the overall evolution of the OC \citep{2013Icar..222...20F,2014Icar..231...99F}.
The effect of planets on the OC is not considered in our simulations and here we discuss how they would contribute to the loss of OC comets.

\citet{2014Icar..231...99F} found that about $12$\% of the OC comets is eventually lost from the OC due to planetary perturbations.
The total OC loss in their models including tides, passing stars, and planetary perturbations is between $24$\% and $41$\%, which is consistent with the lower half of our estimated range ($25$--$65$\%).
However, we do not take planetary perturbations into account and these would increase the OC loss by additional $\sim12$\%.
The discrepancy possibly comes from different treatment of the stellar encounters (the OC initial conditions are the same and the effect of the Galactic tide agrees well).
To derive the distribution of stellar encounters, \citet[][using the same encounters distribution as \citealp{2008CeMDA.102..111R}]{2014Icar..231...99F} used velocity dispersion at the current position of the Sun with mean and standard deviation of about $53$\,\kms{} and $20$\,\kms{}, respectively.
We use the method of \citetalias{2017MNRAS.464.2290M} where the velocity dispersion of the stars is derived from an $N$-body simulation of the Galaxy \citep{bedorf} and changes along the Solar orbit.
For the orbit without any migration, this gives a velocity distribution with a mean of about $46$\,\kms{} and standard deviation of $\sim21$\,\kms{}.
This results in more low velocity encounters that can lead to higher loss from the OC.

\section{Conclusions}
\label{sec:conclusions}

We present numerical simulations of stellar flybys, where the Sun with its OC encounters a field star surrounded by a cloud of exocomets (an exo-OC).
The characteristics of the exo-OCs are scaled with respect to those of the Sun's OC (based on the stellar mass and Galactocentric radius).
We study the possibility that the Sun's Oort cloud contains exocomets that are transferred during the encounters and the erosion of the OC.
In our simulations, the Sun encounters only one star at a time and we considered the effect of the Galactic tide.
We explore the parameter space of such flybys by varying characteristics of the encounter\,---\,the mass of the field star, impact parameter and velocity\,---\,and we measure the rate of the OC capturing exocomets and losing its own comets.
We compare the results with the distributions of encounter characteristics as expected along the sojourn of the Sun through the Galaxy, and we derive the probability that the Sun experienced an encounter resulting in the capture of exocomets to the OC.
Finally we simulate the evolution of the Sun and its OC for $4.5$\,Gyr while accounting for the Galactic tide and encounters with field stars, during which the OC is eroded and it can capture exocomets.

The main findings of our work are as follows:
\begin{enumerate}

\item Considering all flybys of stars passing closer than $5\times10^5$\,AU, we derive that at most $10^{-5}$--$10^{-4}$ of the OC was captured at some point during its evolution.

\item The number of captured exocomets in the OC decreases after the flyby that delivered them.
Similarly to the native comets, exocomets are affected by further stellar encounters and the Galactic tide.

\item The possible radial migration of the Sun does not have a substantial effect on the probability that the Sun's OC captured exocomets. 

\item Over the lifetime of the Solar system, the OC losses between about 25\% and 65\% of its comets due to encounters with field stars.
This loss is dominated by the few strongest encounters along the orbit.

\item We show that the number of exocomets transferred from and to the OC during an encounter is reasonably well approximated by a proxies $\propto \mstar{}^{a}d^{b}v_{\star}^{c}$.
Here the $\mstar{}$ is the mass of the encountered field star (in Solar masses) and $d$ and $v_{\star}$ are the its distance (in AU) and velocity (in \kms{}) with respect to the Sun at infinity and $a$, $b$, and $c$ are free parameters (with different values for comets lost and exocomets captured).

\end{enumerate}

During the work on the manuscript, we became aware of the work of \citet[][see also \citealp{2016DPS....4833103L}]{levine_2017}.
They studied exchange of material between the OC and exo-OCs during flybys of field stars.
Some details of their methods and assumptions are similar to those used here (e.g., every field star posses an exo-OC, the effect of planetary systems is neglected).
Other important details differ (e.g., initial conditions of the clouds, numbers of particles, they do not take the Galactic tide into account), and make a direct comparison with the results presented here difficult.
However, their single star encounter simulations are qualitatively in agreement with those we present here.

\section*{Acknowledgements}

We thank the referee, Ramon Brasser, for a prompt review and useful comments. 
This work was supported by the Netherlands Research School for Astronomy (NOVA) and by the Netherlands Research Council NWO under grant \#621.016.701 (LGM2).
We thank Alan Heays for his comments on the manuscript final version.
LJ acknowledges the hospitality of John Carter and The Institut d'Astrophysique Spatiale (IAS) of the National Center of Scientific Research (CNRS) and of the University of Paris-Sud during the final stages of work on the manuscript.


\bibliographystyle{mnras}
\bibliography{oort_cloud}


\appendix
\section{Comparison of encounters along migrating orbits}
\label{app:encounters_mig}

\begin{figure*}
\center
\begin{tabular}{c}
\includegraphics[width=0.99\textwidth]{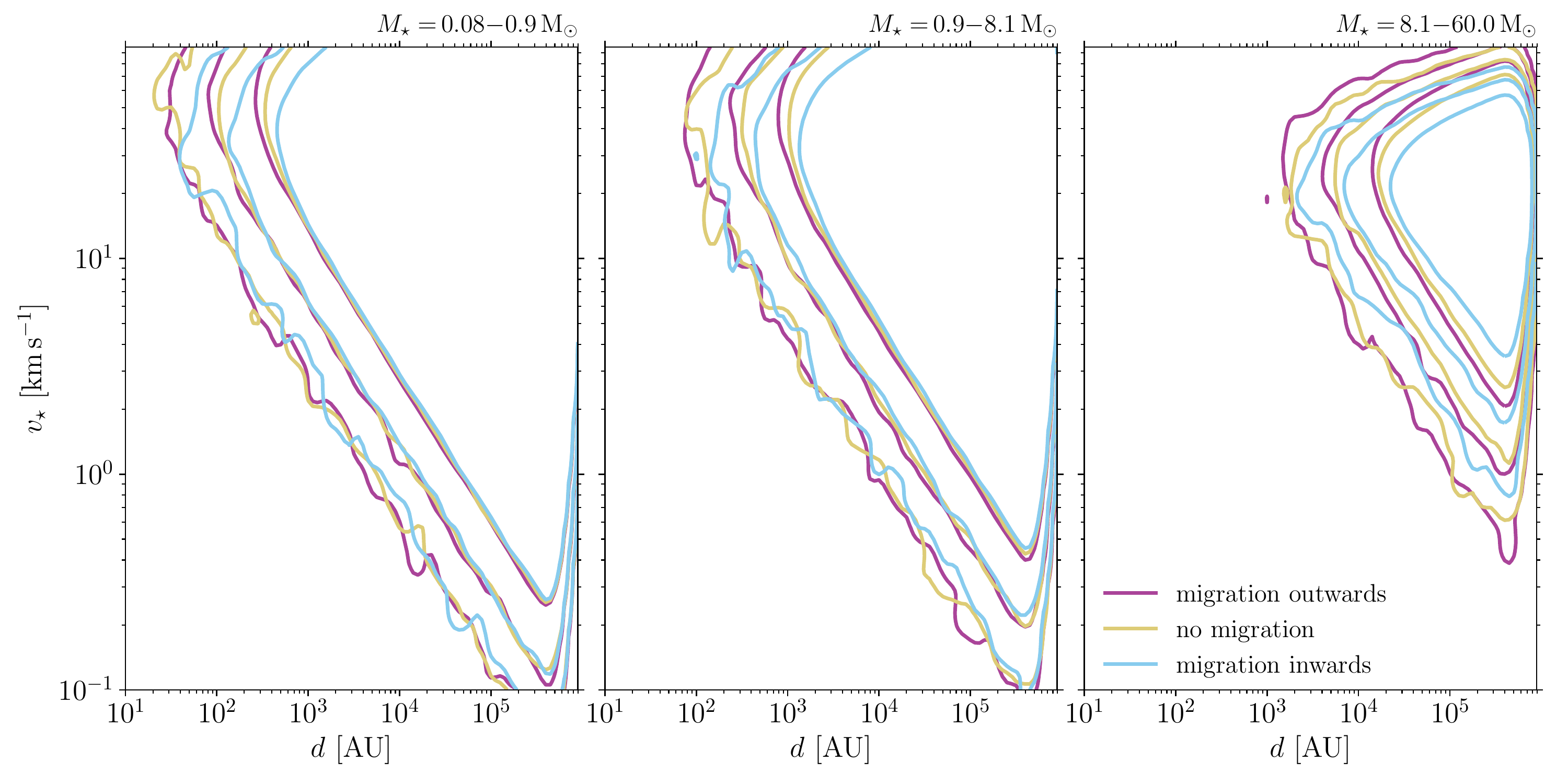}
\end{tabular}
\caption{Density of encounters along the Solar orbit with different migration.
Plots show contours for encounters in three mass bins indicated above each plot.
There are three contours in each plot: outermost, middle and innermost contours correspond to the density of $0.01$, $0.1$, and $1$\,encounter per orbit per $\log (d/\mathrm{AU})\times\log\left(v_{\star}/(\mathrm{km\,s}^{-1})\right)$, respectively.
Lines of different colours show contours for orbits with different migration: purple\,---\,migrating outward, yellow\,---\,no migration (same as contours in Fig.~\ref{fig:mdv}), light blue\,---\,migration inwards.
The contours are averaged over $10^4$ encounter sets.
}
\label{fig:contours_mig}
\end{figure*}

Fig.~\ref{fig:contours_mig} compares the density of encounters along the Solar orbit with different migration (Sec.~\ref{sec:orbitsim} and Fig.~\ref{fig:solar_orbit}).
Three plots of Fig.~\ref{fig:contours_mig} show contours of encounter density for three mass bins ($\mstar{}=0.082$--$0.9$, $0.9$--$8.1$, and $8.1$--$60$\,\msun{}).
The density is normalized by the number of encounters in each mass bin and orbit and by $\log (d/\mathrm{AU})\times\log\left(v_{\star}/(\mathrm{km\,s}^{-1})\right)$.
For each type of orbit, the density is averaged over $10^4$ different encounter sets described in Sec.~\ref{sec:enc}.

\section{Starting position and velocity of encountering star for low \mbox{$v_{\star}$}}
\label{app:vel_ini_for_low_v}

In some cases, for the lowest velocity stars, the method to derive $\mathbfit{r}(t_{\mathrm{enc},0})$ and $\mathbfit{v}(t_{\mathrm{enc},0})$ described in Sec.~\ref{sec:enc_start} does not converge.
The encounter takes too long and there is a strong interaction with the Galactic potential.

In these cases we divide the problem in two parts. 
We start at the current position of the Sun, but instead of integrating for $t_0$, we integrate only for $t_{\mathrm{init}} = 0.5\,t_0$.
From this point onwards we follow the iterative procedure described above.
However this time we do not try to construct an orbit that begins at a separation of $8$\,pc but one that begins at $4$\,pc. 
We now integrate backwards for $t_{\mathrm{init}}$, to get the Sun back to its initial position.
The problem is however that the separation between the Sun and the field star is generally not $8$\,pc now.
We can overcome this by varying $t_{\mathrm{init}}$ until the initial separation is $8$\,pc.
In some cases this procedure does still not converge.
We increase $t_{\mathrm{init}}$ to $0.75\,t_0$, or further, until it does.

In some of the cases in the parameter space study considering the effect of the Galactic tide it is not possible to derive $\mathbfit{r}(t_{\mathrm{enc},0})$ and $\mathbfit{v}(t_{\mathrm{enc},0})$ in the above described way.
$v_{\mathrm{peri}}$ is so small that the distance between Sun and field star never exceeds $8$\,pc due to the interaction with the Galactic potential. 
The Sun and the field star are bound.
This situation occurs in the parameter space study when accounting for the Galactic tide (see Fig.~\ref{fig:mdv_in_potential}), but not in the full orbit simulation.

\section{Grid of single encounters in the Galactic potential}
\label{app:mdv_in_potential}

Here we present the parameter space study (Secs.~\ref{sec:gridsim} and \ref{sec:resultsGridsim}) that takes the Galactic potential into account.
We simulate the encounters with the parameters described in Sec.~\ref{sec:gridsim}.
We use five-times less particles to represent the clouds than in the case without Galactic potential (i.e. $N_{\mathrm{OC}}$ and $N_{\mathrm{exoOC}}$ listed in Table.~\ref{tab:n_mdv}).
The position and velocity of the encounters in the Galaxy is the current position and velocity of the Sun ($\mathbfit{r}_{\odot}=(-8.5,\,0,\,0.02)$\,kpc and $\mathbfit{v}_{\odot}=(-11.1,\, 238.4,\,7.25)$\,km\,s$^{-1}$, see Sec.~\ref{sec:orbitsim}).
We derive the initial position and velocity vectors of the field star with respect to the Sun as described in Sec.~\ref{sec:enc_start} where we further assume that the direction of the vectors are $\mathbfit{d}=(d,0,0)$ and $\mathbfit{v}_{\star}=(0,v_{\star},0)$, respectively.

\begin{figure*}
\center
\includegraphics[width=0.99\textwidth]{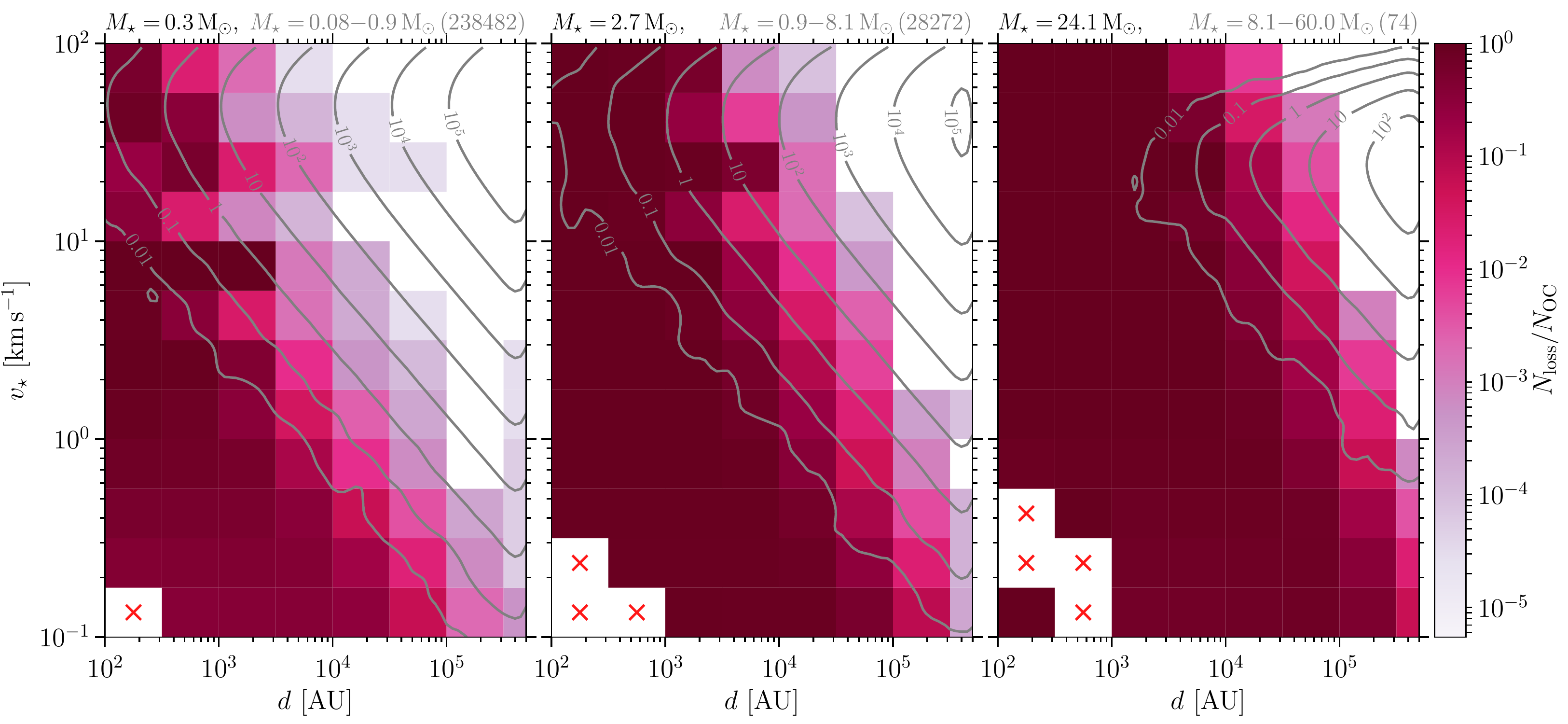} \\
\includegraphics[width=0.99\textwidth]{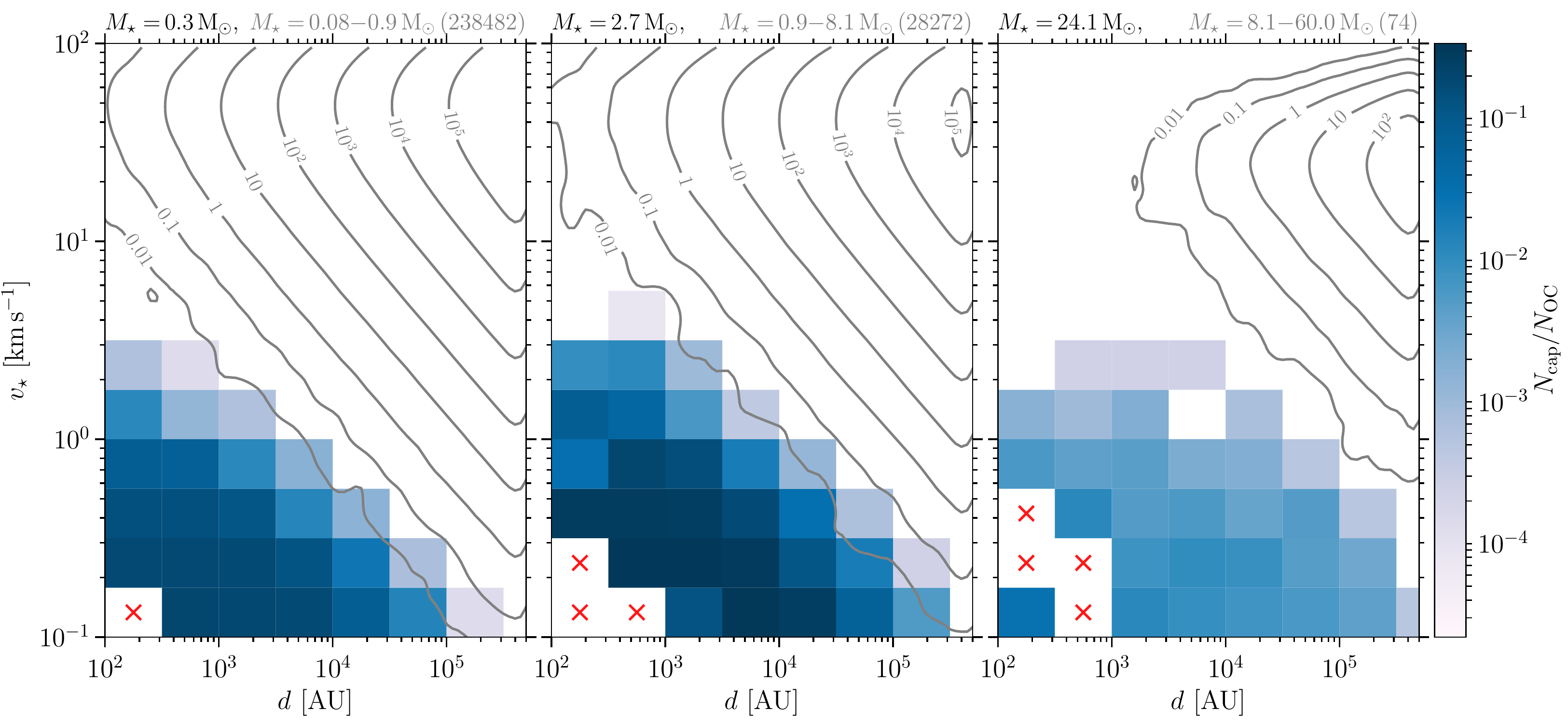}
\caption{Same as Fig.~\ref{fig:mdv} but with the Galactic potential taken into account.
To allow an easier comparison, the colour scales for both $N_{\mathrm{loss}}/\noc{}$ and $N_{\mathrm{cap}}/\noc{}$ are the same as is Fig.~\ref{fig:mdv}.
The red crosses indicate encounters for which the method to derive the initial position and velocity vectors of the encountering star (Sec.~\ref{sec:enc_start}) does not converge.
These are slow ($v_{\star}<0.5$\,\kms{}) and close ($d<10^3$\,AU) encounters that basically do not appear along the Sun's orbit.
}
\label{fig:mdv_in_potential}
\end{figure*}

Fig.~\ref{fig:mdv_in_potential} shows the fraction of the Sun's OC lost and captured during the encounters. 
Comparison with Fig.~\ref{fig:mdv} shows (the colour scale is the same in both figures) that the Galactic potential does not have substantial effect on the results.
For the captured fraction of the OC, the difference of $N_{\mathrm{cap}}/\noc{}$ without and with the effect of the Galactic tide is typically (for more than $85$\% of the bins) $\lesssim 10^{-2}$.
The median difference across all bins (in $M_{\star}$, $v_{\star}$, and $d$) is then $\sim1\times10^{-3}$.
For the lost fraction of the OC, the difference of $N_{\mathrm{loss}}/\noc{}$ without and with the effect of the Galactic tide is typically (for more than $85$\% of the bins) $\lesssim 5\times10^{-2}$ and the median difference across all bins is $\sim3\times10^{-3}$.
However, fractions lost in about $20$ relatively fast encounters with $v_{\star}\gtrsim10$\,\kms{} differ substantially, by $\gtrsim30$\%.
In these cases, the particles that are lost due to Galactic tide, are only weekly bound after the encounter without tide.

\bsp	
\label{lastpage}
\end{document}

%% file: table_compare_mdv_and_sets.tex
\begin{table*}
\caption{Number of encounters along different orbits and average fractions of OC being lost or captured.
The results are given for the lifetime of the Sun and averaged over $1000$ different encounter sets for each type of orbit.
The first column indicates the range of the field star mass ($M_{\star}$).
There are five columns for each orbit (with migration inwards, no migration, and migration outwards):
$N_{\mathrm{enc}}$\,---\,average number of encounters in given mass range $^{a}$;
$N_{\mathrm{enc}}^{\mathrm{cap}}$\,---\,average number of encounters during which exocomets are captured into the OC;
$\eta_{\mathrm{cap}}$\,---\,average captured fraction of the OC (sum of $N_{\mathrm{cap}}/\noc$ resulting from $N_{\mathrm{enc}}^{\mathrm{cap}}$ encounters in all sets divided by $1000$);
$N_{\mathrm{enc}}^{\mathrm{loss}}$\,---\,average number of encounters during which comets are lost from the OC;
$\eta_{\mathrm{loss}}$\,---\,average lost fraction of the OC (see $\eta_{\mathrm{cap}}$);
The last line of the table lists the overall results for all encounters (i.e. $M_{\star}$ from the mass range of $0.082$--$60$\,\msun).
}
\label{tab:n_enc_per_orbit}
\begin{center}
\begin{tabular}{r*{3}{r@{\hspace{0.2cm}}r@{\hspace{0.2cm}}r@{\hspace{0.2cm}}r@{\hspace{0.2cm}}r@{\hspace{0.7cm}}}}
\hline
    & \multicolumn{5}{c}{migration inwards} &
    \multicolumn{5}{c}{no migration} &
    \multicolumn{5}{c}{migration outwards} \\[0.15cm]    
    $M_{\star}$ & 
    $N_{\mathrm{enc}}$ &
    $N_{\mathrm{enc}}^{\mathrm{cap}}$ &
    $\eta_{\mathrm{cap}}$ &
    $N_{\mathrm{enc}}^{\mathrm{loss}}$ &
    $\eta_{\mathrm{loss}}$ &
    $N_{\mathrm{enc}}$ &
    $N_{\mathrm{enc}}^{\mathrm{cap}}$ &
    $\eta_{\mathrm{cap}}$ &
    $N_{\mathrm{enc}}^{\mathrm{loss}}$ &
    $\eta_{\mathrm{loss}}$ &
    $N_{\mathrm{enc}}$ &
    $N_{\mathrm{enc}}^{\mathrm{cap}}$ &
    $\eta_{\mathrm{cap}}$  &
    $N_{\mathrm{enc}}^{\mathrm{loss}}$ &
    $\eta_{\mathrm{loss}}$ \\[0.07cm]
    $[\msun]$ &
    &
    $[10^{-2}]$ &
    $[10^{-5}]$ &
    &
    &
    &
    $[10^{-2}]$ &
    $[10^{-5}]$ &
    &
    &
    &
    $[10^{-2}]$ &
    $[10^{-5}]$ &
    &
    \\[0.15cm]
$0.082$--$0.9$ & 
$1.3\times10^{5}$ & $1.2$ & $ 2.25$ & $ 537$ & $ 0.02$ & $2.4\times10^{5}$ & $1.3$ & $ 4.59$ & $ 787$ & $ 0.02$ & $3.8\times10^{5}$ & $2.1$ & $ 10.65$ & $1043$ & $ 0.03$ \\
$0.9$--$8.1$ &
$ 1.5\times10^{4}$ & $1.2$ & $ 4.13$ & $ 570$ & $ 0.26$ & $ 2.8\times10^{4}$ & $1.9$ & $ 9.36$  & $ 960$ & $ 0.36$ & $ 4.5\times10^{4}$ & $2.3$ & $ 5.65$ & $1394$ & $ 0.49$\\
$8.1$--$60$ &
$    39$ & $0.0$ & $ 0.0$ & $  16$ & $ 0.04$ & $    74$ & $0.1$ & $ 0.09$ & $  30$ & $ 0.07$ & $   118$ & $0.1$ & $ 0.01$ & $  47$ & $ 0.11$  \\
\hline
$0.082$--$60$ &
$1.4\times10^{5}$ & $2.4$ & $ 6.38$ & $1123$ & $ 0.32$ & $2.7\times10^{5}$ & $3.3$ & $14.04$ & $1777$ & $ 0.46$ & $4.2\times10^{5}$ & $4.5$ & $16.31$ & $2484$ & $ 0.63$ \\[0.15cm]
\multicolumn{16}{p{0.99\textwidth}}{$^{a}$\,Note that here the number of encounters is limited by $v_{\star}<10^2\,\kms$ which is the maximum velocity of the bin considered in our grid simulations. 
Therefore the total number of encounters (last line) is about $10^4$ less then the values we mention in Sec.~\ref{sec:enc}.}\\
\end{tabular}
\end{center}
\end{table*}